%% file: main.tex
\documentclass[preprint,3p,12pt]{elsarticle}

\usepackage{amssymb}
\usepackage{booktabs}
\usepackage{nicefrac}
\usepackage{amsmath}
\usepackage{graphicx}
\usepackage{lscape}
\usepackage{multicol}
\usepackage{tabularray}
\usepackage{subcaption}
\usepackage{nicematrix}
\UseTblrLibrary{amsmath}
\usepackage{gensymb}
\makeatletter
\def\maketag@@@#1{\hbox{\m@th\normalfont\normalsize#1}}
\makeatother

\usepackage{diagbox}

\usepackage[colorlinks,bookmarksopen,bookmarksnumbered,citecolor=red,urlcolor=red]{hyperref}

\biboptions{sort&compress}

\journal{Computer Methods in Applied Mechanics and Engineering}

\begin{document}
\input{commands.tex}
\begin{frontmatter}



\title{Structural cohesive element for the modelling of delamination in composite laminates without the cohesive zone limit}


\author[inst1]{Xiaopeng Ai}

\affiliation[inst1]{organization={Department of Aerospace Structures and Materials, Faculty of Aerospace Engineering, Delft University of Technology},
            addressline={Kluyverweg 1}, 
            city={Delft},
            postcode={2629 HS}, 
            country={The Netherlands}}

\author[inst1]{Boyang Chen\corref{cor1}}
 \ead{b.chen-2@tudelft.nl}
 \cortext[cor1]{Corresponding author}
\author[inst1]{Christos Kassapoglou}


\begin{abstract}
Delamination is a critical mode of failure that occurs between plies in a composite laminate. The cohesive element, developed based on the cohesive zone model, is widely used for modeling delamination. However, standard cohesive elements suffer from a well-known limit on the mesh density—the element size must be much smaller than the cohesive zone size. This work develops a new set of elements for modelling composite plies and their interfaces in 3D. A triangular Kirchhoff-Love shell element is developed for orthotropic materials to model the plies. A \emph{structural} cohesive element, conforming to the shell elements of the plies, is developed to model the interface delamination. The proposed method is verified and validated on the classical benchmark problems of Mode I, Mode II, and mixed-mode delamination of unidirectional laminates, as well as on the single-leg bending problem of a multi-directional laminate. All the results show that the element size in the proposed models can be ten times larger than that in the standard cohesive element models, with more than 90\% reduction in CPU time, while retaining prediction accuracy. This would then allow more effective and efficient modeling of delamination in composites without worrying about the cohesive zone limit on the mesh density.
\end{abstract}



\begin{keyword}
Composites \sep delamination \sep cohesive zone model \sep cohesive element
\end{keyword}

\end{frontmatter}


\section{Introduction}
The accurate prediction of delamination is of critical importance for the reliable design of fiber-reinforced composite structures. The cohesive element (CE) is a widely used finite element technology to model delamination. CE is developed based on the Cohesive Zone Model proposed by Dugdale and Barenblatt \cite{barenblatt1962mathematical,dugdale1960yielding}. A fracture process zone, generally called the cohesive zone, exists along the interface, ahead of the stress-free crack tip.  A traction-separation relationship, namely the cohesive law, describes how the interfacial stresses and damage evolve with respect to the interfacial openings. Standard CEs are usually developed for use between two solid elements to model their debonding \cite{qiu2001interface,yang2005cohesive,camanho2002mixed,moes2002extended,xu1993void,turon2007engineering,pirondi2014comparative,chen1999predicting}. 

While being a popular and versatile tool to model delamination, standard CEs suffer from a well-known limit on the mesh density – the element size must be much smaller than the Cohesive Zone Length (CZL) to accurately predict delamination. According to the previous analysis \cite{qiu2001interface,yang2005cohesive}, high-stress gradients could be produced within the cohesive zone during delamination in composites. A very fine CE mesh must be used there to sufficiently capture the stress gradients such that the internal virtual work of the CEs can be accurately integrated. So far, there is no fixed rule on how fine the mesh should be in the literature. In some research \cite{camanho2002mixed,moes2002extended,xu1993void}, the authors have demonstrated that at least two or three CEs should be used inside the cohesive zone. With coarser meshes, simulations would significantly over-predict the peak load \cite{turon2007engineering}. In a typical mode I delamination test of a unidirectional composites coupon, i.e., the Double Cantilever Beam (DCB) test, the CZL is less than 1 mm. 


The above-mentioned problem of cohesive zone limit on the mesh density of CEs has drawn the attention of many researchers in the past. Turon et al. \cite{turon2007engineering} adopted an engineering method to solve the problem by reducing the material strength to numerically extend the CZL. While shown to be promising in the Mode I DCB case, in cases such as the pure mode II End-Notched-Flexure (ENF) test, decreasing the cohesive strength can cause excessive under-prediction of the overall strength \cite{harper2008cohesive,lu2019cohesive}. Yang \cite{yang2010improved} and Do \cite{do2013improved}'s research work demonstrated that larger CE sizes can be achieved by placing enough integration points in the cohesive zone. Their method could predict the peak load correctly for CE size up to 1.43 times the estimated CZL. However, its first-order shape function limits the effectiveness of this method in further expanding CE beyond the scale of the cohesive zone \cite{russo2020overcoming}. Guiamatsia et al.\cite{guiamatsia2009decohesion,guiamatsia2010framework,guiamatsia2013study} used the beam on elastic foundation solution as an enrichment function and tested it in the mixed mode delamination. However, the enrichment method could lead to inaccuracies in interpolation for elements larger than 3 mm \cite{guiamatsia2010framework}. Another enrichment approach with the piecewise linear shape functions was proposed by Samimi \cite{samimi2009enriched,samimi2011self,samimi2011three}. However, the over-prediction problem of peak load under large CE has not been solved. Van der Meer et al. \cite{van2012level} used the level set method with a energy-based criterion to propagate delamination without the cohesive zone limit. However, this method is limited to the case of a single delamination. Lu et al. \cite{lu2018adaptive} proposed a adaptive version of the Floating Node Method (FNM) \cite{chen2014floating} to adaptively refine the CEs in the cohesive zone. However, this method is currently implemented for 2D problems. Alvarez et al. \cite{alvarez2014mode} used quadratic CEs between quadratic solid elements and used higher number of integration points, which increased the grid size to be comparable to CZL but not beyond \cite{russo2020overcoming}. Mukhopadhyay and Bhatia \cite{mukhopadhyay2023accurate} developed a $hp$ refinement strategy to simulate the delamination between two solid elements. However, it is only implemented in 2D in their work. Daniel \cite{daniel2023efficient} developed an ERR-Cohesive method to simulate the delamination with large elements by estimating the energy release rate by means of the virtual crack closure technique (VCCT). However, this method is only presented in 2D, and the VCCT approach relies on the existence of an initial crack and the assumption of self-similar crack propagation. 


Inspired by the earlier works in the literature, Russo and Chen\cite{russo2020overcoming} developed a so-called structural CE, which conforms to Kirchhoff-Love structural elements for the neighbouring plies. Their work was done in 2D. The Euler-Bernoulli beam elements were used to model the plies. The structural CE, sharing its nodes with the beam elements, was developed to model the delamination.  An adaptive integration scheme was used to place more integration points in CEs containing the cohesive zone. Their results showed that the structural CE could overcome the cohesive zone limit on mesh density, allowing the element size to be ten times larger than that of the standard linear CEs. Motivated by Russo and Chen’s work, Tosti Balducci and Chen \cite{TostiBalducci2024overcoming} extended the structural CE to 3D DCB problem by developing a structural CE compatible with the TUBA3 plate elements \cite{bell1969refined}. Their results showed that the TUBA3-based structural formulation of CE could overcome the cohesive zone limit. However, the curvature degrees of freedom (DoFs) make it more complicated to set boundary conditions on TUBA3 elements than on more common elements such as solid, beam or Reissner-Mindlin shell elements, thereby impeding the adoption of such elements by engineers in practice. 

From the above reviews, we can see that a composites delamination model which: 
\begin{enumerate}
    \item does not suffer from the cohesive zone limit on mesh density;
    \item works in 3D space;
    \item does not require artificial reduction of strengths;
    \item does not require a pre-crack and the self-similar propagation assumption;
    \item can model delamination along multiple interfaces; and
    \item does not require working with curvature DoFs,
\end{enumerate}
is still missing. Based on the earlier works \cite{russo2020overcoming,TostiBalducci2024overcoming}, this article aims to develop a new set of elements for the plies and interfaces of composites. The idea is to develop a triangular Kirchhoff-Love shell element without curvature DoFs to model the composite plies, then formulate the conforming structural CE to be used between the plies to model their delamination.  Allman \cite{allman1976simple} proposed a simple triangle Kirchhoff-Love plate element named cubic displacement element. There are only three DoFs at each node, one for out-of-plane displacement and two for rotations. These DoFs are commonly used by engineers and are much easier to handle than the curvature ones when setting boundary conditions. Therefore, this article will extend this triangle plate element for the modeling of composite plies and develop the corresponding structural CE for the interfaces. If such a structural CE could overcome the cohesive zone limit on its mesh density, then it would achieve all the requirements listed above.

The rest of the paper will be structured in the following way. Section~\ref{sec: method} presents the proposed element formulations in detail. Section~\ref{sec:results} demonstrates the performance of these elements on a series of benchmarks on delamination in both unidirectional and multi-directional laminates. In the end, Section~\ref{sec:conclusions} draws the conclusions of this work and discusses some potential future work.

\section{Method}\label{sec: method}
\subsection{Overall illustration}

\begin{figure}[h]
      \centering
	   \begin{subfigure}{0.4\linewidth}
		\includegraphics[width=\linewidth]{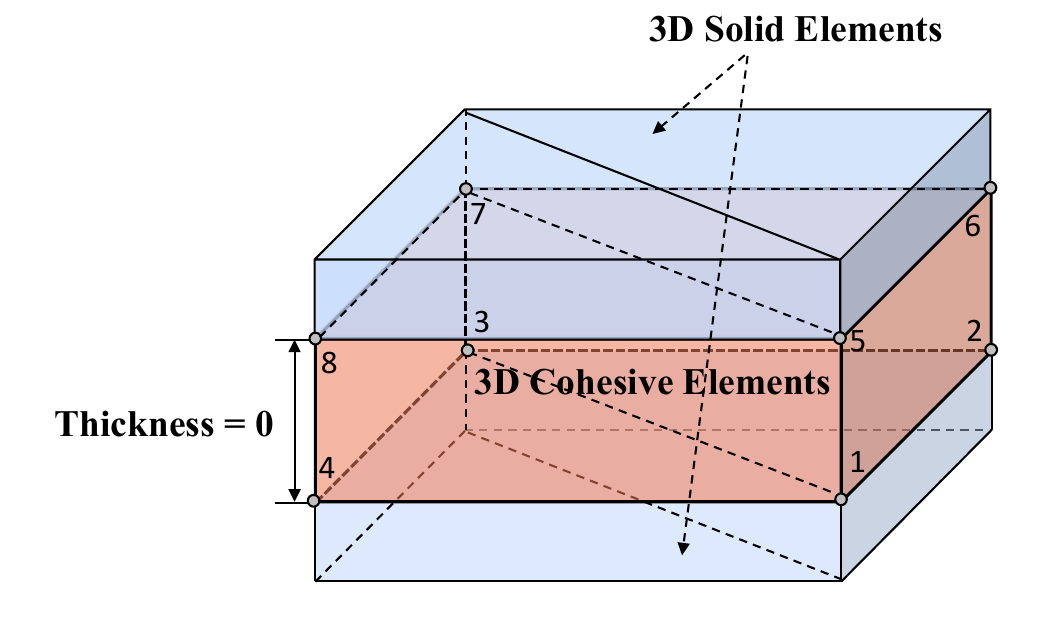}
		\caption{Conventional CE between solids: interface thickness = distance between node 4 and node 8}
		\label{fig:CE_solid&shell-subfig1}
	   \end{subfigure}
    ~
	   \begin{subfigure}{0.4\linewidth}
		\includegraphics[width=\linewidth]{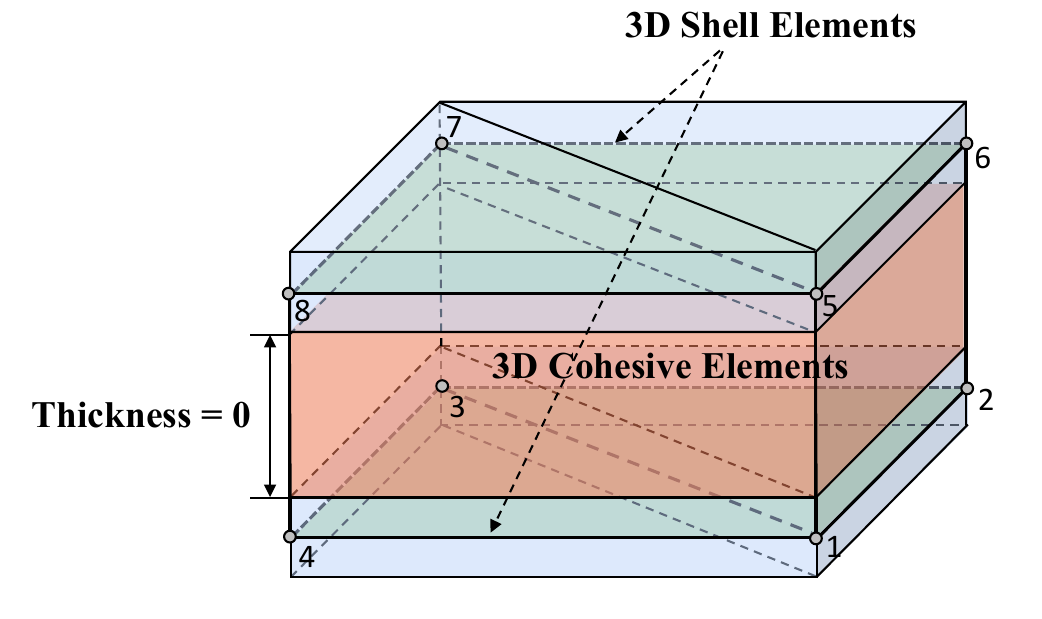}
		\caption{Structural CE between shells: interface thickness $\ne$ distance between node 4 and node 8}
		\label{fig:CE_solid&shell-subfig2}
	    \end{subfigure}
    \caption{Geometrical comparison between conventional CE and structural CE} 
	\label{fig:CE_solid&shell}
\end{figure}

The overall idea of the proposed modelling approach is to represent the composite plies by Kirchoff-Love shell elements and the interfaces by structural CEs. Figure~\ref{fig:CE_solid&shell} shows the geometrical comparison between the conventional modelling approach with solid elements and the proposed structural approach with shells. Looking at the structural CE, the difference from the conventional CE is that the nodes of the CE are placed at the mid-plane of the two shell elements, which are not the actual surfaces of the cohesive interface. Hence, the opening of the actual interface does not equate to the distance between the upper nodes and lower ones of the structural CE. This opening shall be calculated using the shell kinematics to be detailed in later subsections. The rest of this section will firstly present the cubic plate element formulation by Allman \cite{allman1976simple} and its adaptation for flat composites shell, then move on to derive the structural CE formulation.


\subsection{Cubic plate element}\label{subsec: plate element}

\subsubsection{Geometric definitions of the plate element}
\begin{figure}[h]
	\centering
	\includegraphics[width=0.7\linewidth]{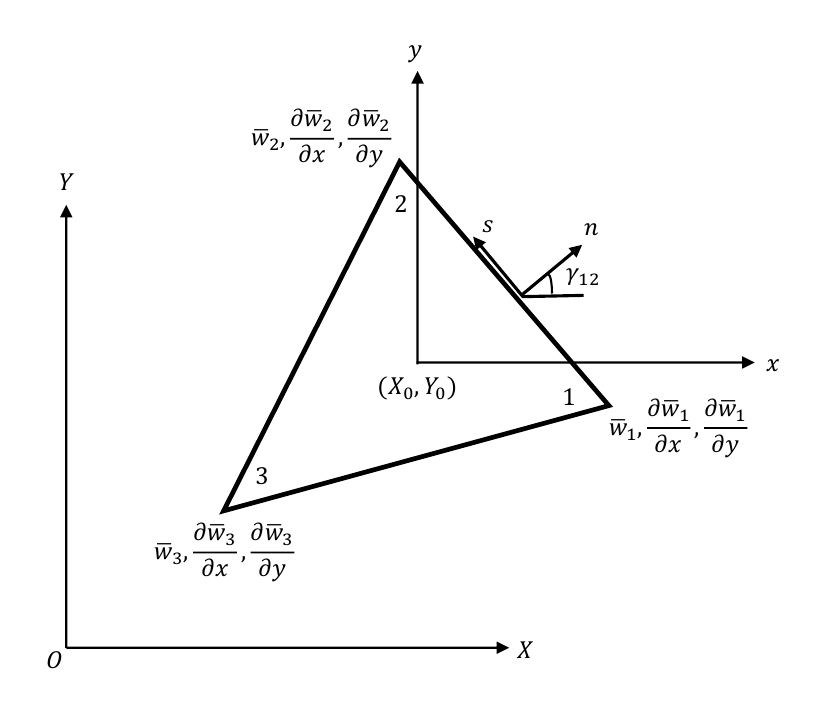}
	\caption{Coordinate system and DoFs for the triangular cubic plate element \cite{allman1976simple}}
	\label{fig: local}
\end{figure}

The plate part of the shell element in this work is based on the triangular cubic plate element developed by Allman \cite{allman1976simple}, as shown in Figure~\ref{fig: local}. A local coordinate system is used in this element, with origin at the centroid of the triangle element. The local axis are represented by lowercase letters, $x$ and $y$, to distinguish them from the global ones, $X$ and $Y$. The area of this triangular element is $A$. $s$ is the anti-clockwise coordinate along the element boundary. $n$ is the exterior normal. The angle between the normal $n$ and the local axis $x$ is $\gamma$. The out-of-plane displacement defined over the domain is ${w}(x, y)$. An independent out-of-plane boundary displacement, $\overline{w}(s)$, and its compatible normal derivative, $\nicefrac{\partial \overline{w}}{\partial n} (s)$, are assumed along the boundary. The DoFs at each node include the displacement $\overline{w}$ and two rotations $\nicefrac{\partial \overline{w}}{\partial x}$, $\nicefrac{\partial \overline{w}}{\partial y}$. 

The cubic plate element in Allman's work only considered the case of isotropic material \cite{allman1976simple}. This work extends the cubic plate element using the classical laminate theory, such that symmetric composite laminates can also be modelled by this plate element. The original cubic element formulation will be presented in detail, with adaptations for composites specified along the way.

\subsubsection{Modified potential energy for the plate element}\label{sec:MPE}



The minimum potential energy principle is used to derive the finite element formulation of the triangular plate element. The potential energy used is referred to as the modified potential energy in Allman's work \cite{allman1976simple}:
  \begin{align}\label{eq:modified PE}
    \pi = \iint_{A} U_0 \dd x \dd y + \sum_{N=1}^{3} R_N(\overline{w}_N -w_N) + \int_{\partial A} V_n(\overline{w} -w) \, \dd s - \int_{\partial A} M_n(\frac{\partial \overline{w}}{\partial n} - \frac{\partial w}{\partial n})\dd s \nonumber\\[0.5em]
    - \iint_{A} p^*w  \dd x\dd y - \sum_{N=1}^{3} R^*_N \overline{w}_N -\int_{\partial A} V^*_n \overline{w} \dd s +  \int_{\partial A} M^*_n \frac{\partial \overline{w}}{\partial n} \, \dd s 
\end{align}    
where $U_0$ is the strain energy density. $M_n$ is the normal bending moment resultant and $V_n $ is the Kirchhoff shear force distribution on the element boundary, respectively. $R_{N \, (N=1,2,3)}$ are the concentrated forces at the element vertices. In addition, $R^*_N$, $V^*_n$, and $ M^*_n$ are the values of prescribed concentrated force, Kirchhoff shear force, and normal bending moment resultant, respectively. $p^*(x,y)$ denotes the prescribed distributed pressure load on the element. This potential energy essentially uses $R_N$, $V_n$, and $ M_n$ as Lagrange multipliers to enforce the compatibility between the two fields $w$ and $\overline{w}$ and their normal derivatives along the boundary. The variational principle based on this modified potential energy satisfies the equilibrium equations, the boundary conditions, and the compatibility requirement between $w$ and $\overline{w}$ \cite{allman1976simple}.

The expression of $U_0$ for composites will be different from that for isotropic materials in Allman's work \cite{allman1976simple}. The classical laminate theory can be used to describe the constitutive relationship of a composite laminate under Kirchhoff-Love kinematic assumptions. For the case of a symmetric laminate (the smallest of which would be a single composite ply) considered in this work, there would be no membrane-bending coupling. Therefore, the expressions of the moments are:
\begin{align}\label{eq:MXMYMXY}
    M_x = - D_{11} \frac{\partial^2 w}{\partial x^2}  - D_{12} \frac{\partial^2 w}{\partial y^2} - 2D_{16} \frac{\partial^2 w}{\partial x \partial y}\nonumber \\[0.5em]
    M_y = - D_{12} \frac{\partial^2 w}{\partial x^2}  - D_{22} \frac{\partial^2 w}{\partial y^2} - 2D_{26} \frac{\partial^2 w}{\partial x \partial y} \nonumber\\[0.5em]
    M_{xy} = - D_{16} \frac{\partial^2 w}{\partial x^2}  - D_{26} \frac{\partial^2 w}{\partial y^2} - 2D_{66} \frac{\partial^2 w}{\partial x \partial y} 
\end{align}

Using the above moment-curvature relations (equation \ref{eq:MXMYMXY}), the strain energy density $U_0$ can be written as:
\begin{align}\label{eq:U0}
    U_0 &= \frac{1}{2} \left[D_{11} \left(\frac{\partial^2 w}{\partial x^2} \right)^2  + 
    D_{22} \left( \frac{\partial^2 w}{\partial y^2} \right)^2 + 2D_{12} \frac{\partial^2 w}{\partial x^2}\frac{\partial^2 w}{\partial y^2}\right.\nonumber \\[0.5em]
    &\left. + 4D_{16}\frac{\partial^2 w}{\partial x^2} \frac{\partial^2 w}{\partial x \partial y} + 4D_{26}\frac{\partial^2 w}{\partial y^2}\frac{\partial^2 w}{\partial x \partial y} + 4D_{66}\left(\frac{\partial^2 w}{\partial x \partial y}\right)^2 \right]
\end{align}

\begin{figure}[h]
    \centering
    \includegraphics[width=0.7\linewidth]{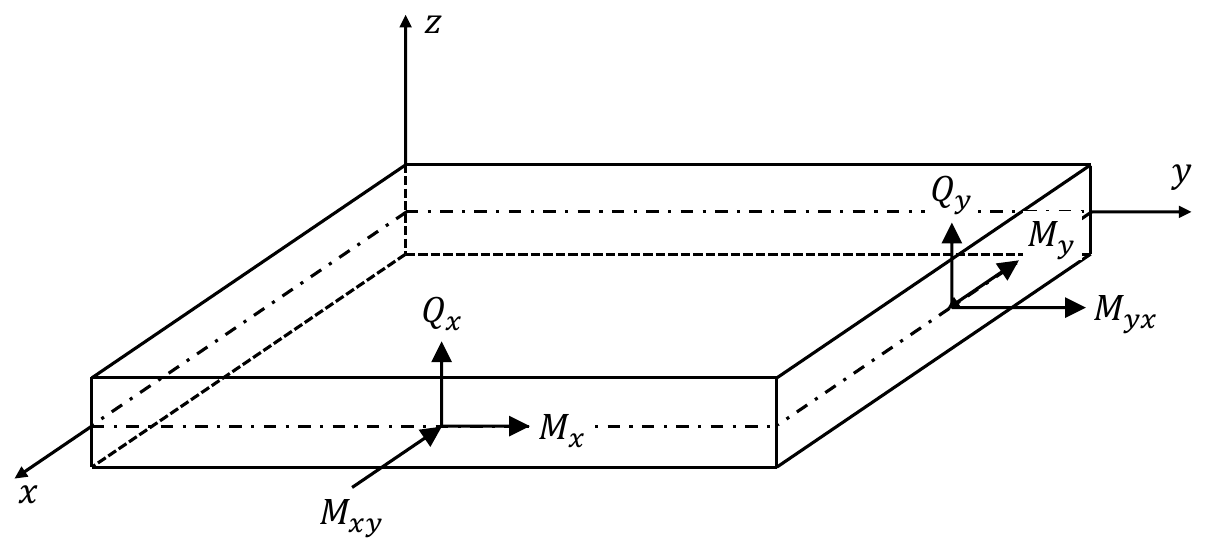}
    \caption{Sign conventions for shear force and bending moment resultants}
    \label{fig:plate_mq}
\end{figure}

The sign conventions for the shear force and bending moment resultants are shown in Figure \ref{fig:plate_mq}. For the triangular plate element, the normal bending moment resultant $M_n$, Kirchhoff shear force distribution $V_n $ and concentrated forces $R_{N \, (N=1,2,3)}$ can be expressed as \cite{allman1976simple}:
\begin{align}
    M_n &= M_x \cos^2{\gamma} + M_y \sin^2{\gamma} + M_{xy} \sin{2\gamma} \label{eq:moment1-1}\\[0.5em]
    V_n &= Q_n + \frac{\partial M_{ns}}{\partial s} \label{eq:moment1-2}\\[0.5em]
    R_N &= [M_{ns}]^{s^+}_{s^-} \label{eq:moment1-3}
\end{align}
where
\begin{align}
    &M_{ns} = \frac{1}{2}(M_y-M_x)\sin 2\gamma + M_{xy} \cos 2\gamma\label{eq:moment2-1}\\[0.5em]
    &Q_n    = \frac{\partial M_n}{\partial n} + \frac{\partial M_{ns}}{\partial s} \label{eq:moment2-2}
\end{align}

In this element, linear variation of the normal bending moment resultant and constant Kirchhoff shear force distribution are assumed along the edges:
\begin{align}
    M_n &= M^{12}_n(1-\xi) + M^{21}_n\xi \label{eq: Mn&Vn-Mn}\\[0.5em]
    V_n &= V_n^{12} \label{eq: Mn&Vn-Vn}
\end{align}
where $M_n^{12}$ and $M_n^{21}$ are the normal bending moment resultant values at node $1$ and node $2$ of side 1-2 respectively. The coefficient $\xi$ in equation \ref{eq: Mn&Vn-Mn} is defined as $\xi = s/l_{12}$. $V_n^{12}$ is the value of the Kirchhoff shear force distribution along the side 1-2. The cyclic permutation of superscripts 1, 2, and 3 in equations \ref{eq: Mn&Vn-Mn} and \ref{eq: Mn&Vn-Vn} produces the shear force distribution and moment resultant on the other two sides (side 2-3 and side 3-1) of the triangular element. 

The terms in brackets in equation \ref{eq:moment1-3} represent the difference in bending moment resultant values at the element vertices. Therefore, the expressions of $R_{N \, (N = 1, 2, 3)}$ are:
\begin{align}\label{eq:R1,R2,R3}
    R_1 = M_{ns}^{12} - M_{ns}^{13} \nonumber \\[0.5em] 
    R_2 = M_{ns}^{23} - M_{ns}^{21}  \nonumber\\[0.5em]   
    R_3 = M_{ns}^{31} - M_{ns}^{32}    
\end{align}

Finally, the directional derivatives in equations \ref{eq:moment2-2} are given by:
\begin{align}\label{eq:directional derivatives}
    &\frac{\partial}{\partial s} = - \sin\gamma \frac{\partial}{\partial x} + \cos \gamma \frac{\partial}{\partial y} \nonumber\\[0.5em]
    &\frac{\partial}{\partial n} =  \cos\gamma \frac{\partial}{\partial x} + \sin \gamma \frac{\partial}{\partial y}
\end{align}

\subsubsection{Finite element approximation and the discretized equilibrium equation}\label{sec:3 matrices}
For finite element implementation, the potential energy $\pi$ should be discretized. In this element, the internal out-of-plane displacement field $w(x, y)$ is approximated by a cubic polynomial:
\begin{equation}\label{eq:cubicw}
    w(x, y) = A_1 + A_2\, x + A_3\,y + \alpha_1 \, x^2 + \alpha_2 \, xy + \alpha_3 \, y^2 + \alpha_4 \, x^3 + \alpha_5 \, x^2 y + \alpha_6 \, x y^2 + \alpha_7 \, y^3
\end{equation}
The three coefficients $A_1$, $A_2$, and $A_3$ represent the rigid body motion and do not affect the value of the strain energy density.

With the above choice of $w$, applying the Green's theorem \cite{przemieniecki1985theory} to transform the strain energy integral to boundary integral gives:
\begin{align}\label{eq: Green's theorem}
    2\iint_{A} U_0 \, \dd x \dd y = \sum_{N=1}^{3} R_Nw_N + \int_{\partial A} V_nw \, \dd s - \int_{\partial A} M_n \frac{\partial w}{\partial n}\, \dd s   
\end{align}  
In this work, we assume that there is no distributed pressure load $p^*(x,y)$. Then, the modified potential energy (equation \ref{eq:modified PE}) can be rewritten more simply as:
\begin{align}\label{eq:simple modified PE}
    \pi = -\iint_{A} U_0 \, \dd x \dd y + \sum_{N=1}^{3} R_N\overline{w}_N + \int_{\partial A} V_n\overline{w} \, \dd s - \int_{\partial A} M_n\frac{\partial \overline{w}}{\partial n} \, \dd s \nonumber\\
    - \sum_{N=1}^{3} R^*_N \overline{w}_N -\int_{\partial A} V^*_n \overline{w}\, \dd s +  \int_{\partial A} M^*_n \frac{\partial \overline{w}}{\partial n} \, \dd s 
\end{align}    

Substituting equation \ref{eq:cubicw} into \ref{eq:U0}, the strain energy can be rewritten as:
\begin{equation}\label{eq:U0=0.5aHa}
    \iint_{A} U_0 \, \dd x \dd y = \frac{1}{2} \balpha \transpose \, \bH \, \balpha
\end{equation}
where the vector $\balpha$ is:
\begin{equation}\label{eq:vector alpha}
    \balpha = \{\alpha_1,\, \alpha_2, \, \alpha_3, \, \alpha_4, \, \alpha_5, \, \alpha_6, \, \alpha_7 \}\transpose
\end{equation}
The matrix $\bH$ is more complicated to derive for symmetric composite laminates than for isotropic materials considered in reference~\cite{allman1976simple}. Its detailed derivation is presented in \ref{sec:H-matrix}.

After the strain energy, the $2^\mathrm{nd}$, $3^\mathrm{rd}$ and $4^\mathrm{th}$ terms in equation \ref{eq:simple modified PE} denote the work of the so-called generalized forces:
\begin{align}\label{eq:pi_234}
    \sum_{N=1}^{3} R_N\overline{w}_N + \int_{\partial A} V_n\overline{w} \, \dd s - \int_{\partial A} M_n\frac{\partial \overline{w}}{\partial n} \, \dd s = \bQ^{\rm{T}}\bq
\end{align}
where $\bQ$ is the vector of the twelve generalized forces:
\begin{equation}\label{eq:bQ}
    \bQ = \{R_1, R_2, R_3, V_n^{12},V_n^{23},V_n^{31},M_n^{12},M_n^{21},M_n^{23},M_n^{32},M_n^{31},M_n^{13}\}\transpose
\end{equation}
Substituting equations \ref{eq: Mn&Vn-Mn} and \ref{eq: Mn&Vn-Vn} into equation \ref{eq:pi_234}, the generalized displacements corresponding to the generalized forces in equation \ref{eq:bQ} compose the vector $\bq$:
\begin{align}\label{eq:vector_q}
    \bq = &\left\{\overline{w}_1,\,\overline{w}_2,\,\overline{w}_3,\;l_{12}\int_0^1 \overline{w} \, \dd \xi,\; l_{23}\int_0^1 \overline{w} \, \dd \xi,\; l_{31}\int_0^1 \overline{w} \, \dd \xi, \right.\nonumber\\
    &\left. -l_{12}\int_0^1 \frac{\partial\overline{w} }{\partial n}(1-\xi) \, \dd \xi, \; -l_{12}\int_0^1 \frac{\partial\overline{w} }{\partial n}\xi \, \dd \xi, \; -l_{23}\int_0^1 \frac{\partial\overline{w} }{\partial n}(1-\xi) \, \dd \xi, \right.\nonumber\\
    &\left. -l_{23}\int_0^1 \frac{\partial\overline{w} }{\partial n}\xi \, \dd \xi, \; -l_{31}\int_0^1 \frac{\partial\overline{w} }{\partial n}(1-\xi) \, \dd \xi, \; -l_{31}\int_0^1 \frac{\partial\overline{w} }{\partial n}\xi \, \dd \xi \right\}\transpose
\end{align}

A matrix $\bB$ can be constructed  to connect the vector $\bQ$ and the vector $\balpha$:
\begin{equation}\label{eq:Q=Ba}
    \bQ = \bB \transpose \balpha
\end{equation}
With the moment-curvature relationship for composites in equation~\ref{eq:MXMYMXY} and the expression of $w$ in equaton~\ref{eq:cubicw}, the matrix $\bB$ has been derived for composites in this work. The details are shown in \ref{sec:B-matrix}.

A matrix $\bT$ can be defined to  represent the relationship between the nodal DoFs $\overline{\bW}$ and the vector $\bq$:
\begin{align}\label{eq:q=TW}
    \bq = \bT \overline{\bW}
\end{align}
where $\overline{\bW}$ is the vector of the nodal DoFs defined by
\begin{align}
    \overline{\bW} =\left\{ \overline{w}_1,\frac{\partial \overline{w}_1}{\partial x},\frac{\partial \overline{w}_1}{\partial y}, 
    \overline{w}_2,\frac{\partial \overline{w}_2}{\partial x},\frac{\partial \overline{w}_3}{\partial y}, 
    \overline{w}_3,\frac{\partial \overline{w}_3}{\partial x},\frac{\partial \overline{w}_3}{\partial y} \right\}\transpose
\end{align}
With the assumptions of cubic line function for $\overline{w}$ and linear variation for $\nicefrac{\partial\overline{w}}{\partial n}$ along each edge of the element, the $\bT$ matrix can derived as \cite{allman1976simple}:
\begin{scriptsize}
\begin{equation}\label{eq:T matrix}
    \begin{+bmatrix}[
      colspec={@{~}ccccccccc@{~}},
      colsep=2.2pt
    ]
    1 & 0 & 0 & 0 & 0 & 0 & 0 & 0 & 0 \\[1em]
    0 & 0 & 0 & 1 & 0 & 0 & 0 & 0 & 0 \\[1em]
    0 & 0 & 0 & 0 & 0 & 0 & 1 & 0 & 0 \\[1em]
    \dfrac{l_{12}}{2} & \minus \dfrac{l^2_{12}}{12} \sin{\gamma_{12}} &  \dfrac{l^2_{12}}{12} \cos{\gamma_{12}} & \dfrac{l_{12}}{2} & \dfrac{l^2_{12}}{12} \sin{\gamma_{12}} & \minus\dfrac{l^2_{12}}{12} \cos{\gamma_{12}} & 0 & 0 & 0  \\
    0 & 0 & 0 & \dfrac{l_{23}}{2} & \minus\dfrac{l^2_{23}}{12} \sin{\gamma_{23}} &  \dfrac{l^2_{23}}{12} \cos{\gamma_{23}} & \dfrac{l_{23}}{2} & \dfrac{l^2_{23}}{12}\sin{\gamma_{23}} & \minus\dfrac{l^2_{23}}{12} \cos{\gamma_{23}}\\
    \dfrac{l_{31}}{2} &  \dfrac{l^2_{31}}{12} \sin{\gamma_{31}} & \minus\dfrac{l^2_{31}}{12} \cos{\gamma_{31}} & 0 & 0 & 0 & \dfrac{l_{31}}{2} & \minus\dfrac{l^2_{31}}{12} \sin{\gamma_{31}} & \dfrac{l^2_{31}}{12} \cos{\gamma_{31}} \\
    0 & \minus\dfrac{l_{12}}{3} \cos{\gamma_{12}} & \minus\dfrac{l_{12}}{3} \sin{\gamma_{12}} & 0 &  \minus\dfrac{l_{12}}{6} \cos{\gamma_{12}} & \minus\dfrac{l_{12}}{6} \sin{\gamma_{12}} & 0 & 0 & 0 \\
    0 & \minus\dfrac{l_{12}}{6} \cos{\gamma_{12}} & \minus\dfrac{l_{12}}{6} \sin{\gamma_{12}} & 0 &  \minus\dfrac{l_{12}}{3} \cos{\gamma_{12}} & \minus\dfrac{l_{12}}{3} \sin{\gamma_{12}} & 0 & 0 & 0 \\
    0 & 0 & 0 & 0 & \minus\dfrac{l_{23}}{3} \cos{\gamma_{23}} & \minus\dfrac{l_{23}}{3} \sin{\gamma_{23}} & 0 & \minus\dfrac{l_{23}}{6} \cos{\gamma_{23}} & \minus\dfrac{l_{23}}{6}  \sin{\gamma_{23}} \\
     0 & 0 & 0 & 0 & \minus\dfrac{l_{23}}{6} \cos{\gamma_{23}} & \minus\dfrac{l_{23}}{6} \sin{\gamma_{23}} & 0 & \minus\dfrac{l_{23}}{3} \cos{\gamma_{23}} & \minus\dfrac{l_{23}}{3} \sin{\gamma_{23}} \\
     0 & \minus\dfrac{l_{31}}{6} \cos{\gamma_{31}} & \minus\dfrac{l_{31}}{6} \sin{\gamma_{31}} & 0 & 0 & 0 & 0 & \minus\dfrac{l_{31}}{3} \cos{\gamma_{31}} &  \minus\dfrac{l_{31}}{3} \sin{\gamma_{31}} \\
     0 & \minus\dfrac{l_{31}}{3} \cos{\gamma_{31}} & \minus\dfrac{l_{31}}{3} \sin{\gamma_{31}} & 0 & 0 & 0 & 0 & \minus\dfrac{l_{31}}{6} \cos{\gamma_{31}} &  \minus\dfrac{l_{31}}{6} \sin{\gamma_{31}}
    \end{+bmatrix}
\end{equation}
\end{scriptsize}


\subsubsection{Stiffness matrix and force vectors of cubic plate element}\label{sec:Kmatrix_plate}
Substituting equations \ref{eq:U0=0.5aHa}, \ref{eq:Q=Ba} and \ref{eq:q=TW} into
equation \ref{eq:simple modified PE}, the total modified potential energy represented by the finite element method under prescribed boundary loads is:
\begin{align}\label{eq:PI}
    \pi = - \frac{1}{2}\balpha\transpose\bH\balpha + \balpha\transpose(\bB\bT)\overline{\bW}  - \bQ^{*\rm{T}}\bT\overline{\bW}
\end{align}
where the vector $\bQ^*$, whose components are obtained by replacing the generalized forces in vector $\bQ$ with the corresponding prescribed quantities, denotes all the external forces:
\begin{align}
    \bQ^* = \{R_1^*, R_2^*, R_3^*, V_n^{*12},V_n^{*23},V_n^{*31}, M_n^{*12},M_n^{*21},M_n^{*23},M_n^{*32},M_n^{*31},M_n^{*13}\}\transpose
\end{align}

Thus, the internal work $U$ and the external work $W$ of the potential energy can be expressed as:
\begin{align}
    U &= - \frac{1}{2}\balpha\transpose\bH\balpha + \balpha\transpose(\bB\bT)\overline{\bW} \label{eq:U-discrete}\\[0.5em]
    W &= {\bQ^*}\transpose\bT\overline{\bW}
\end{align}

Based on equation~\ref{eq:PI}, the minimum total potential energy principle gives:
\begin{align}
    \delta \pi = \delta \balpha\transpose[(\bB\bT)\overline{\bW}-\bH\balpha] + \delta\overline{\bW}\transpose[(\bB\bT)\transpose\balpha - \bT\transpose \bQ^*] = 0
\end{align}
Setting the coefficient of the arbitrary variation $\delta\balpha\transpose$ to zero gives:
\begin{align}\label{eq:alpha}
    \balpha = \bH^{-1}(\bB\bT)\overline{\bW}
\end{align}

Performing the variation of $U$ as expressed in \ref{eq:U-discrete} with $\balpha$ substituted by equation~\ref{eq:alpha}, we arrive at $\delta U$ for the element as:
\begin{align}
    \delta U &= \delta\overline{\bW}\transpose(\bB\bT)\transpose\bH^{-1}(\bB\bT)\overline{\bW}\label{eq:deltaU-1} = \delta \overline{\bW}\transpose \fint 
\end{align}
where $\fint$ is the internal force vector of the plate element:
\begin{align}
   \fint = (\bB\bT)\transpose \bH^{-1} (\bB\bT) \, \overline{\bW} = \K_\mathrm{plate} \, \overline{\bW}
\end{align}
and $\K_\mathrm{plate}$ is the stiffness matrix of the plate element:
\begin{align}\label{eq:Kmatrix_plate}
    \K_\mathrm{plate} = (\bB\bT)\transpose \bH^{-1} (\bB\bT)
\end{align}

The variation of the external work $W$ gives the external force vector of the element in the absence of distributed pressure load:
\begin{align}
    \delta W = \delta \overline{\bW} \, \bT\transpose \bQ^* = \delta \overline{\bW} \, \fext, \quad \Rightarrow \quad \fext = \bT\transpose \bQ^*
\end{align}


\subsection{Formation of shell element} \label{subsec: shell element}
To form a flat shell element that considers membrane deformation and bending, we superimpose a linear membrane element on top of the plate element developed in the previous section. The displacement fields $u$, $v$ are defined in terms of area coordinates $L_1$, $L_2$, $L_3$:
\begin{align}\label{eq:uv}
    u &= u_1 L_1 + u_2 L_2 + u_3 L_3 \\[0.5em]
    v &= v_1 L_1 + v_2 L_2 + v_3 L_3 
\end{align}
where $u_i$ and $v_i$ represent the displacements along $x$ and $y$ for node $i$, respectively. The DoFs vector of the membrane element $\overline{\bU}$ is:
\begin{align}
    \overline{\bU} = \left\{u_1, \, v_1, \, u_2, \,v_2, \,u_3, \,v_3\right\}\transpose
\end{align}

The stiffness matrix of the triangular membrane element is given by the expression:
\begin{align}\label{eq:kmem}
    \K_\mathrm{mem}= \iint_\Delta \bB_\mathrm{mem}\transpose \,\D_\mathrm{mem}\, \bB_\mathrm{mem} \, t \,\dd x \,\dd y 
\end{align}
where $t$ is the thickness of the membrane element and $\bB_\mathrm{mem}$ matrix in equation \ref{eq:kmem} is
\begin{align}
    \bB_\mathrm{mem} =  \frac{1}{2A} 
    \begin{bmatrix}
    b_1 & 0 & b_2 & 0 & b_3 & 0 \\[0.5em]
    0 & c_1 & 0 & c_2 & 0 & c_3 \\[0.5em]
    c_1 & b_1 & c_2 & b_2 & c_3 & b_3 
    \end{bmatrix}
\end{align}
$A$ is the area of the membrane element and: 
\begin{align}
    b_1 &= y_2 - y_3, \quad
    b_2 = y_3 - y_1, \quad
    b_3 = y_1 - y_2 \\
    c_1 &= x_3 - x_2, \quad
    c_2 = x_1 - x_3, \quad
    c_3 = x_2 - x_1
\end{align}

The $\D^\mathrm{mem}$ matrix in equation \ref{eq:kmem} is:
\begin{align}\label{eq:Dmat_MEM}
    \D_\mathrm{mem}
 = 
    \begin{bmatrix} 
    \frac{E_1}{1-\nu_{12}\nu_{21}} \,& \frac{\nu_{12}E_2}{1-\nu_{12}\nu_{21}} \, & 0 \\[1.2em]  \,& \frac{E_2}{1-\nu_{12}\nu_{21}}\, & 0 \\[1.2em]  \text{symmetric} \,& \, & G_{12} 
    \end{bmatrix} 
\end{align}

From equations~\ref{eq:Kmatrix_plate} and \ref{eq:kmem}, we have obtained the stiffness matrices of the plate and membrane elements, respectively. They are assembled to form the stiffness matrix of the shell element: 
\begin{align}\label{eq:Kmatrix_shell}
    \K_\mathrm{shell} =     \begin{bmatrix} 
     \K_\mathrm{mem}\, & 0 \\[1.5em]   0\, & \K_\mathrm{plate}
    \end{bmatrix} 
\end{align}
The corresponding DoF vector of the shell element is simply:
\begin{align}\label{eq: dof of shell}
    \bq_\mathrm{shell} &=\Bigl\{\overline{\bU}, \overline{\bW}\Bigl\} \\\nonumber &= \Bigl\{  u_1, v_1, u_2, v_2, u_3, v_3, {w}_1,\frac{\partial \overline{w}_1}{\partial x},\frac{\partial \overline{w}_1}{\partial y},
    \overline{w}_2,\frac{\partial \overline{w}_2}{\partial x},\frac{\partial \overline{w}_3}{\partial y}, 
    \overline{w}_3,\frac{\partial \overline{w}_3}{\partial x},\frac{\partial \overline{w}_3}{\partial y} \Bigl\}\transpose
\end{align}

The contribution of the shell element to the overall force residual of the finite element system equation is $\fext - \fint$. When the applied loads are nodal forces, $\fext$ does not need to be calculated explicitly in the element subroutine as the nodal forces can be directly entered in the global external force vector of the system. In this case, the residual contribution of this element can be expressed as:
\begin{align}
 \f_{\mathrm{res}} = - \fint = -\K_{\text{shell}} \, \bq_\mathrm{shell}
\end{align}



\subsection{Structural CE between two plies}\label{subsec: cubic-ce}
\subsubsection{DoFs and opening vector}
The structural CEs must be compatible with the top and bottom shell elements kinematically. Thus, they need to share the same DoFs and displacement interpolations along the interface. The DoF vector of the CE can be defined as:
\begin{align}\label{eq:qCE}
    \bq_{\CE} &= \left[\overline{\bU}_{\text{bot}}\transpose, \overline{\bW}_{\text{bot}}\transpose, \overline{\bU}_{\text{top}}\transpose, \overline{\bW}_{\text{top}}\transpose \right] \transpose
\end{align}
where $\overline{\bU}_{\text{bot/top}}, \overline{\bW}_{\text{bot/top}}$ are the membrane and plate DoFs of the bottom/top ply, respectively. 

\begin{figure}[h]
	\centering
	\includegraphics[width=0.85\linewidth]{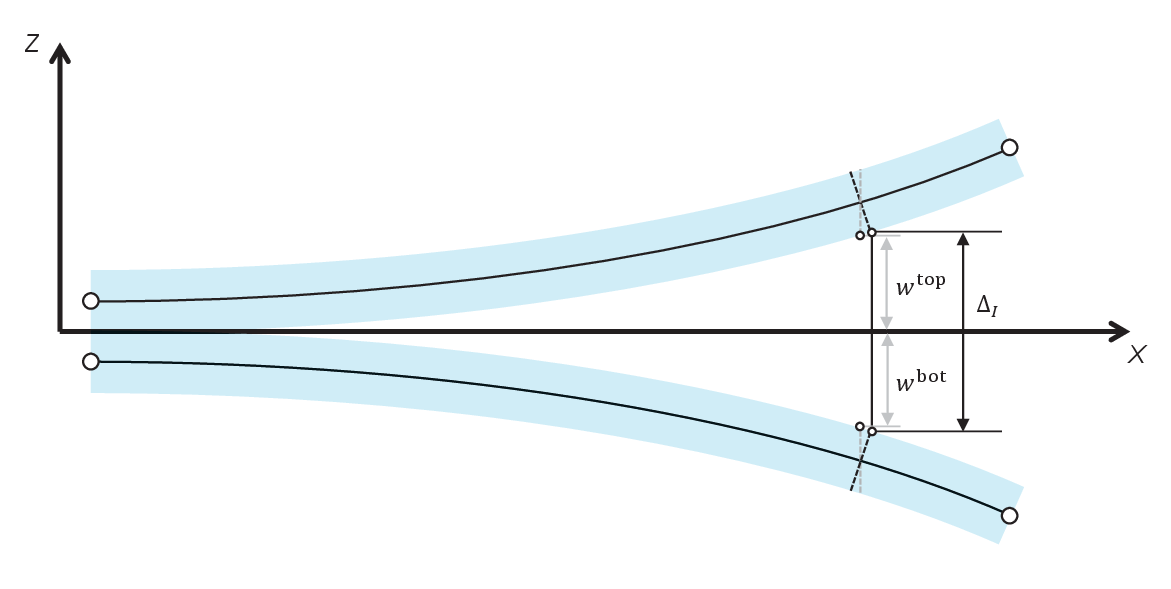}
	\caption{Mode I opening}
	\label{fig:Mode1}
\end{figure}

The Mode I opening displacement of the CE can be expressed as
\begin{align}
    \DeltaI = \wtopce - \wbotce
\end{align}
where $w^{\text{topCE}}$ and $w^{\text{botCE}}$ are the vertical displacements of the top and bottom CE surfaces, respectively. From Figure \ref{fig:Mode1}, using classical plate theory, we have:
\begin{align}
     w^{\text{topCE}}  = w^{\text{top}} + \frac{\htop}{2} (1-\cos{\thetatop}) \\[0.5em]
w^{\text{botCE}}  = w^{\text{bot}} - \frac{\hbot}{2} (1-\cos{\thetabot}) 
\end{align}
$\htop$ and $\hbot$ are the thickness of top and bottom plies, respectively. $w^{\text{top}}$ and $w^{\text{bot}}$ are the vertical displacements of the neutral planes of top and bottom plies, respectively. In this work, only geometrical linearity is considered. Thus, the rotations are small such that $1-\cos{\theta} \approx 0$. The mode I opening is then simply the relative displacement between the mid-planes of the plies:
\begin{align}\label{eq:delta_I=wt-wb}
    \DeltaI = \wtop - \wbot 
\end{align}

\begin{figure}[h]
	\centering
	\includegraphics[width=0.85\linewidth]{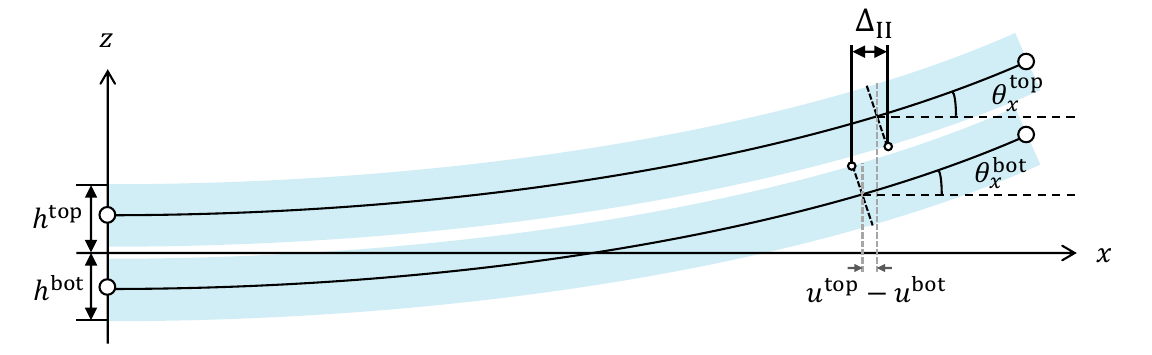}
	\caption{Mode II opening}
	\label{fig:Mode2}
\end{figure}

The Mode II opening of the CE shown in Figure \ref{fig:Mode2} can be expressed as:
\begin{align}
    \DeltaII = \utopce - \ubotce 
\end{align}
where $\utopce$ and $\ubotce$ are the displacements along the x-axis of the top and bottom CE surfaces, respectively. Considering the rotations of the shells’ neutral planes and the offsets of shell surfaces from the neutral planes, they can be written as:
\begin{align}
    u^{\text{topCE}}  = u^{\text{top}} + \frac{\htop}{2} \sin{\thetatop_x} \\[0.5em]
u^{\text{botCE}}  = u^{\text{bot}} - \frac{\hbot}{2} \sin{\thetabot_x}
\end{align}
where $u^{\text{top}}$, $u^{\text{bot}}$ are the displacement $u$ of the neutral planes of the top and bottom shells, respectively. For a small rotation $\theta$ in geometrically linear problems, $\sin{\theta} \approx \theta$. Then, the opening in mode II is:
\begin{align}\label{eq:delta_II}
    \DeltaII = \utop - \ubot + \frac{\htop}{2} \thetatop_x + \frac{\hbot}{2} \thetabot_x
\end{align}

The mode III opening is found from analogous kinematics in the yz-plane. Hence:
\begin{align}\label{eq:delta_III}
    \DeltaIII = \vttop - \vbot + \frac{\htop}{2} \thetatop_y + \frac{\hbot}{2} \thetabot_y
\end{align}

The central task of CE formulation is to find the matrix, $\bB_{\CE}$, that relates the opening vector to its nodal DoFs:
\begin{align}\label{eq:Bce}
    \bDelta = \left[ \DeltaI, \DeltaII, \DeltaIII \right]\transpose = \bB_{\CE} \, \bq_{\CE}
\end{align}
Examining the expressions of $\DeltaI, \DeltaII, \DeltaIII$ in equations~\ref{eq:delta_I=wt-wb}, \ref{eq:delta_II}, and \ref{eq:delta_III}, we can see that the $\bB_{\CE}$ matrix shall be composed of sub-matrices that relate the following terms to the nodal DoFs in $\bq_{\CE}$:
\begin{align}
    \wtop, \, \wbot, \, \utop, \, \ubot, \, \vttop, \, \vbot, \, \thetatop_x, \, \thetabot_x, \, \thetatop_y, \, \thetabot_y
\end{align}
In equation~\ref{eq:uv}, the in-plane displacements $u$ and $v$ are already expressed in the nodal membrane DoFs. However, the out-of-plane displacements $w$ and the rotations $\theta$ in the above list remain to be explicitly expressed in terms of nodal DoFs in $\bq_{\CE}$. 

\subsubsection{Shape functions of $w$}\label{sec:Expressions of w}
The displacement $w$ shall be expressed in terms of plate DoFs $\overline{\bW}$ in this section. Referring back to equation \ref{eq:cubicw}, the displacement $w$ is defined in the local Cartesian coordinates ($x$ and $y$) by the coefficients $A_i$ and the vector $\balpha$. The expression of $w$ is then divided into two parts. The first part only contains $A_i$, and the second part only contains $\alpha_j$:
\begin{align}\label{eq:w=wA+w_a}
  w = w_A + w_{\alpha}
\end{align}
where
\begin{align}
   w_A &= A_1 + A_2\, x + A_3\,y \\[0.5em]
   w_{\alpha} &= \alpha_1 \, x^2 + \alpha_2 \, xy + \alpha_3 \, y^2 + \alpha_4 \, x^3 + \alpha_5 \, x^2 y + \alpha_6 \, x y^2 + \alpha_7 \, y^3 
\end{align}
$w_{A}$ can be rewritten as:
\begin{align}\label{eq:w_A=STA}
    w_{A} = \bS\transpose \, \bA, \quad \bS = [1, \, x, \, y] \transpose, \; \bA = [A_1, \, A_2, \, A_3]\transpose
\end{align}
Similarly, $w_{\alpha}$ can be rewritten as:
\begin{align}\label{eq:walpha=ra}
    w_{\alpha} = \bR\transpose \, \balpha, \quad \bR = [x^2, \, xy, \,y^2, \,x^3 , \,x^2 y, \,x y^2, \,y^3] \transpose
\end{align}

Let us firstly look at $w_{\alpha}$. Based on equation \ref{eq:alpha}, $\balpha$ and $\overline{\bW}$ can be related by multiplying the three matrices $\bH$, $\bB$ and $\bT$ in section \ref{sec:Kmatrix_plate}. Here, for simplicity, we replace the product of these three matrices with matrix $\bC$:
\begin{align}
    \bC =  \bH^{-1}(\bB\bT)
\end{align}
Thus, the vector $\balpha$ can be rewritten as:
\begin{align}\label{eq:alpha_appendix}
\balpha =  \bC\overline{\bW}
\end{align}
Substituting equation \ref{eq:alpha_appendix} in to \ref{eq:walpha=ra}, we can obtain the expression of $w_{\alpha}$ in terms of $\overline{\bW}$:
\begin{align}\label{eq:w_a_final}
    w_{\alpha} = \bR\transpose \bC\overline{\bW} 
\end{align}



Next, we move on to express $w_{A}$ in terms of $\overline{\bW}$. From equations~\ref{eq:w=wA+w_a}, \ref{eq:w_A=STA} and \ref{eq:w_a_final}, we have:
\begin{align}
    w_{A}= \bS\transpose \, \bA = w - \bR\transpose \bC\overline{\bW}
\end{align}
Evaluating $w_{A}$ at the three nodes, we can obtain:
\begin{align}
    w_{A}(x_i,y_i)=\bS\transpose(x_i,y_i) \, \bA =\overline{w}_i - \bR\transpose(x_i,y_i) \, \bC \overline{\bW}, \quad i=1, 2, 3
\end{align}
which give us the following matrix equation:
\begin{align}
  \underbrace{\begin{bmatrix} 
  1 \,& x_1\, & y_1 \\[0.5em] 1 \,& x_2\, & y_2 \\[0.5em]  1 \,& x_3\, & y_3
  \end{bmatrix}}_{\bM_A} \bA = \begin{bmatrix}
  \overline{w}_{1} \\[0.5em] \overline{w}_{2} \\[0.5em] \overline{w}_{3}  
\end{bmatrix} - \underbrace{\begin{bmatrix}
  x_1^2 & x_1y_1 & y_1^2 & x_1^3 & x_1^2y_1 & x_1y_1^2  & y_1^3  \\[0.5em]
  x_2^2 & x_2y_2 & y_2^2 & x_2^3 & x_2^2y_2 & x_2y_2^2  & y_2^3  \\[0.5em]
  x_3^2 & x_3y_3 & y_3^2 & x_3^3 & x_3^23_2 & x_3y_3^2  & y_3^3 
\end{bmatrix}}_{\bM_{\alpha}} \bC \overline{\bW}
\end{align}
Since
\begin{align}
\begin{bmatrix}
  \overline{w}_{1} \\[0.5em] \overline{w}_{2} \\[0.5em] \overline{w}_{3}  
\end{bmatrix}=
     \underbrace{\begin{bmatrix}
      1 & 0 & 0 & 0 & 0 & 0 & 0 & 0 & 0 \\[0.5em]
      0 & 0 & 0 & 1 & 0 & 0 & 0 & 0 & 0 \\[0.5em]
      0 & 0 & 0 & 0 & 0 & 0 & 1 & 0 & 0 
    \end{bmatrix}}_{\bB_A} \overline{\bW}
\end{align}
we can obtain $\bA$ as:
\begin{align}\label{eq:A_by_W}
    \bA = \bM_A^{-1} \left(\bB_A - \bM_\alpha \, \bC\right) \overline{\bW}
\end{align}
Using equation~\ref{eq:w_A=STA}, we obtain:
\begin{align}
    w_{A}= \bS\transpose \bM_A^{-1} \left(\bB_A - \bM_\alpha \, \bC\right) \overline{\bW}
\end{align}
Combining with equation~\ref{eq:w_a_final}, we obtain the expression of $w$:
\begin{align}\label{eq:w_by Nw}
    w &= \bS\transpose \bM_A^{-1} \left(\bB_A - \bM_\alpha \, \bC\right) \overline{\bW} + \bR\transpose \bC\overline{\bW} = \bN_{w} \, \overline{\bW}
\end{align}
where the matrix $\bN_{w}$ ($1\times 9$) can be extracted as: 
\begin{align}\label{eq:N_w}
   \bN_{w} = \bS\transpose \bM_A^{-1}(\bB_A -  \bM_\alpha \bC) + \bR\transpose \bC 
\end{align}

\subsubsection{Shape functions of $\theta$}

Using the cubic polynomial expression of $w$ from equation~\ref{eq:cubicw}, $\theta_x$ can be expressed as:
\begin{align}
\theta_x = \frac{\partial w}{ \partial x} = A_2 + 2\alpha_1 x + \alpha_2 y + 3\alpha_4 x^2 + 2\alpha_5 xy + \alpha_6 y^2 \label{eq:thetay_extend}
\end{align}

Similarly as in the case of $w$, the expression of the above rotations can also be divided into two parts, one containing the $A$ coefficients only and another the $\alpha$ coefficients only: 
\begin{align}
    \theta_{x} = A_2 + \theta_{x\alpha}
\end{align}
Since $A_2$ is part of the vector $\bA$ which has been expressed in equation~\ref{eq:A_by_W}, we have:
\begin{align}
    A_2 = \bB_x \bA = \bB_x \bM_A^{-1}(\bB_A -  \bM_{\alpha} \bC) \overline{\bW}
\end{align}
where $\bB_x$ is a Boolean matrix:
\begin{align}
    \bB_x = \begin{bmatrix}
      0 & 1 & 0
    \end{bmatrix}
\end{align}

The expression $\theta_{x\alpha}$ can be written as:
\begin{align}
    \theta_{x\alpha} = \bR_x\transpose \balpha = \bR_x\transpose  \bC\overline{\bW} 
\end{align}
where equation~\ref{eq:alpha_appendix} is used to express $\balpha$ and $\bR_x$ is also a vector of local coordinates: 
\begin{align}
    \bR_x = \begin{bmatrix}
      2x & y & 0 & 3x^2 & 2xy & y^2 & 0
    \end{bmatrix} \transpose
\end{align}
Substituting the expressions above into equation \ref{eq:thetay_extend}, the expression of $\theta_{x}$ is
\begin{align}
    \theta_{x} &= \bB_x \bM_A^{-1}(\bB_A -  \bM_{\alpha} \bC) \overline{\bW} + \bR_x\transpose \bC\overline{\bW} 
\end{align}
The shape function $\bN_x$ of rotation $\theta_x$ is therefore:
\begin{align}\label{eq:N_x}
   \bN_x = \bB_x \bM_A^{-1}(\bB_A -  \bM_{\alpha} \bC) + \bR_x\transpose \bC
\end{align}

The derivation of shape function $\bN_y$ for rotation $\theta_y$ is very similar to that for $\theta_x$ and is omitted here for brevity.

\subsubsection{$\bB_{\CE}$ matrix}
Once we have determined the shape functions, the next step is to assemble the $\bB_{\CE}$ matrix that relates the DoFs vector $\bq_{\CE}$ to the opening vector $\bDelta$ (c.f. equation~\ref{eq:Bce}). 
Substituting the shape functions in equations~\ref{eq:N_w} and \ref{eq:N_x} to express the $w$ and $\theta$ terms in equations~\ref{eq:delta_I=wt-wb}, \ref{eq:delta_II}, and \ref{eq:delta_III}, the expression of $\bB_{\CE}$ can be obtained as the following: 

      \begin{align}
      \bB_{\CE}=
     \begin{+bmatrix}[
      colspec={@{~}cccccccccccccc@{~}},
      colsep=3pt
    ]
      0 & 0 & 0 & 0 & 0 & 0 & -\bN_w & 0 & 0 & 0 & 0 & 0 & 0 & \bN_w \\[1em]
      -L_1 & 0 & -L_2  & 0 & -L_3  & 0 & \frac{\hbot}{2}\bN_x & L_1 & 0 & L_2  & 0 & L_3  & 0 & \frac{\htop}{2} \bN_x \\[1em]
       0 & -L_1 & 0 & -L_2  & 0 & -L_3 & \frac{\hbot}{2}\bN_y & 0 & L_1 & 0 & L_2  & 0 & L_3 & \frac{\htop}{2}\bN_y
    \end{+bmatrix}
\end{align}  

\subsubsection{$\D_{{\CE}}$ matrix}
The relationship between traction $\btau$ and opening vector $\bDelta$ is expressed through the constitutive matrix $\D_{\CE}$:
\begin{align}\label{eq:Dmat_coh}
\btau = \D_{\CE} \, \bDelta, \quad
  \D_{\CE} = 
\begin{bmatrix}
  (1-d_\text{I})K & 0 & 0 \\[0.5em]
    0      & (1-d)K & 0 \\[0.5em]
    0      &    0   & (1-d)K  
\end{bmatrix}  
\end{align}
where $K$ and $d$ are the penalty stiffness and the damage variable of the CE, respectively. The damage variable $d$ in this work is updated by the bi-linear cohesive law proposed by Turon et al. \cite{turon2006damage,turon2010accurate}. The damage variable $ d_\text{I} $ under Mode I loading is distinguished from $ d $ to avoid interpenetration of the top and bottom surfaces under compression:
\begin{align}
  d_\text{I} = \begin{cases}\label{eq:damagevar_I}
  d,\quad  &  \Delta_\text{I} \geq 0 \\[0.5em]
  0,\quad  & \Delta_\text{I} < 0
  \end{cases}
\end{align}
The penalty stiffness $K$ is set as:
\begin{align}
  K = \alpha\frac{E_3}{t}
\end{align} 
where $E_3$ is the out-of-plane laminate Young's modulus, $t$ is the thickness of the laminate and $\alpha$ is a constant, here set to be 50 \cite{turon2007engineering}.

\subsubsection{Stiffness matrix and residual vector}
In this work, the secant stiffness matrix is used as the stiffness matrix of the CE:
\begin{align}\label{eq:kmat_coh}
 \K_{\CE} = \int_{\Gamma}  \bB_{\CE}\transpose \, \D_{\CE} \, \bB_{\CE} \, \dd \, \Gamma 
\end{align}
where $\Gamma$ represents the domain of the interface. The integral in equation \ref{eq:kmat_coh} is hard to calculate analytically. Thus, the Gaussian integration scheme is applied to obtain the stiffness matrix numerically. Earlier works have shown that using a higher number of quadrature points improves the accuracy and smoothness of the load-displacement solutions of delamination simulations \cite{alvarez2014mode,russo2020overcoming,TostiBalducci2024overcoming}. Therefore, thirteen quadrature points are used in this work for the integration of the structural CE, with their coordinates and weights taken from the work of Cowper \cite{cowper1973gaussian}. 

Assuming that no external distributed loads are applied to the cohesive interfaces, only the internal force vector of the CE contributes to the overall residual vector of the system. The residual vector contribution from this CE can then be written as:
\begin{align}
  \f_{\mathrm{res}} = - \fint = -\K_{\CE} \, \bq_{\CE}
\end{align}

\section{Results}
\label{sec:results}
The structural elements section~\ref{sec: method} have been implemented in the Abaqus user-defined element subroutines. The structural CE model was firstly verified on the three classical benchmarks, namely the double cantilever beam (DCB), the end-notched flexure (ENF), and the mixed-mode bending (MMB) problems. As the above benchmarks only pertain to unidirectional laminates, a multi-directional laminate problem, namely the single-leg bending (SLB) problem \cite{davidson1995three,krueger2015summary}, was simulated to demonstrate the 3D capacity of the model. All simulations of the structural CE model were performed with the Quasi-Newton solver without damping or viscosity. The reference solutions on these problems were obtained with analytical equations and Abaqus standard FE simulations using 8-node linear solid elements (C3D8I element in Abaqus) for the plies and 8-node linear CEs for the interfaces. The linear CEs were implemented as user-defined elements following the formulation in \cite{camanho2002mixed}, with the cohesive law from Turon et al. \cite{turon2006damage,turon2010accurate}. The problem descriptions and modelling details will be presented in this section, followed by comparisons of load-displacement curves and computational time. 

\subsection{Unidirectional laminate benchmarks: DCB, ENF, and MMB}
\subsubsection{Description of the unidirectional laminate tests}
 
\begin{figure}[h!]
       \centering
	   \begin{subfigure}{0.48\linewidth}
		\includegraphics[width=\linewidth]{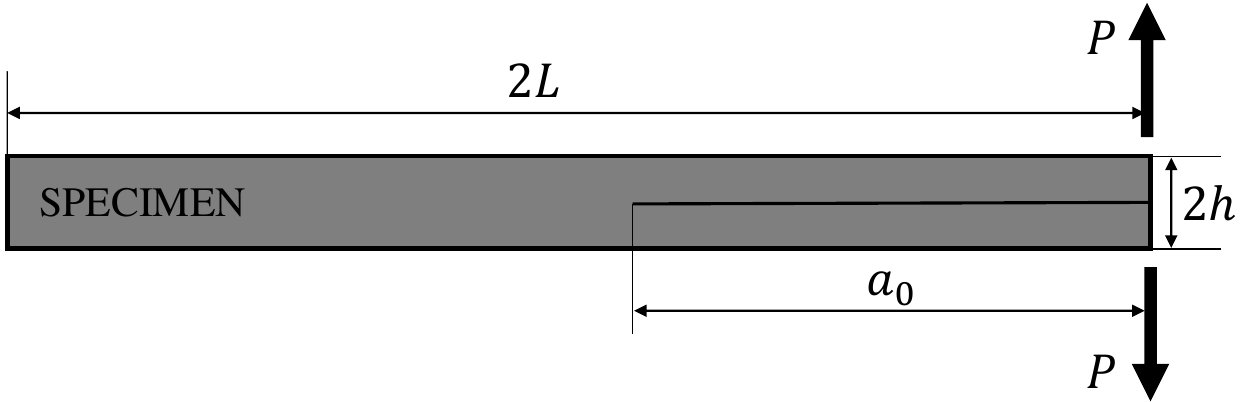}
		\caption{DCB}
		\label{fig:dcb}
	   \end{subfigure}
    ~
	     \begin{subfigure}{0.48\linewidth}
		 \includegraphics[width=\linewidth]{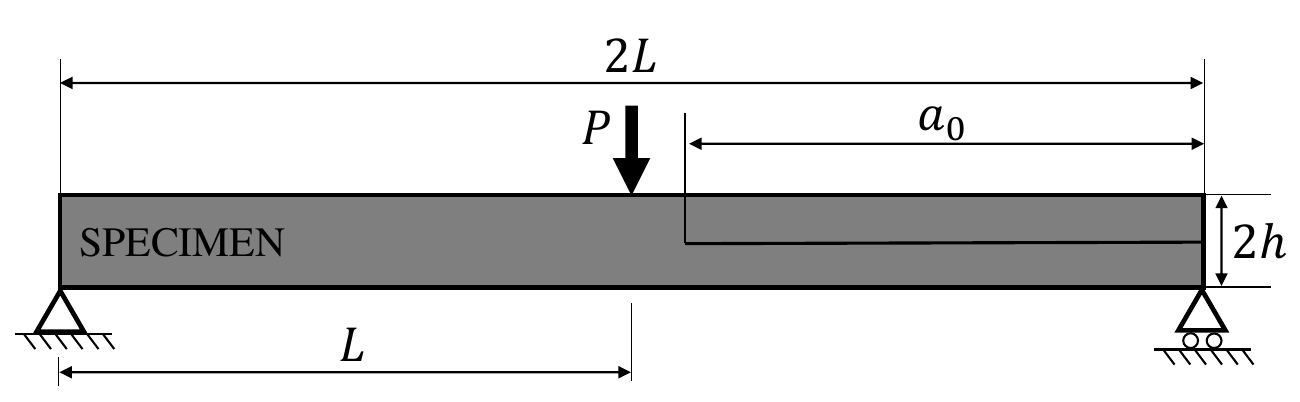}
		 \caption{ENF}
		 \label{fig:enf}
	      \end{subfigure}
      \vfill  
      \begin{subfigure}{0.48\linewidth}
		 \includegraphics[width=\linewidth]{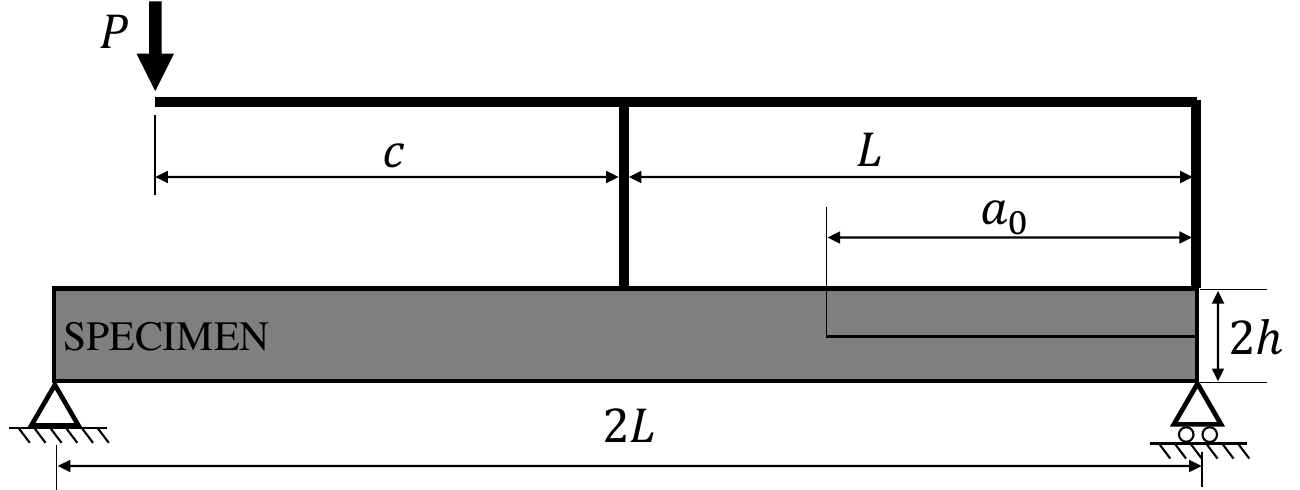}
		 \caption{MMB}
		 \label{fig:mmb}
	      \end{subfigure}
	\caption{DCB, ENF, and MMB test specimens}
	\label{fig:UDbenchmarks}
\end{figure}

The unidirectional laminate benchmarks are drawn from the work by Krueger \cite{krueger2015summary}, except that the pre-crack length of the ENF model is increased to 35 mm in this work to avoid the snapback in load-displacement response. The geometric parameters and boundary conditions are shown in Figure \ref{fig:UDbenchmarks} and Table \ref{tab:DCBgeo}. The detailed material properties are shown in Table \ref{tab:T300Material} and \ref{tab:ENFmaterial}.

\begin{table}[h!]
  \centering
  \caption{Geometric parameters for unidirectional benchmarks}
    \begin{tabular}{l|ccclc}
    \hline
    \backslashbox{Model}{Param. (mm)}
    &\makebox[3em]{$2L$}&\makebox[3em]{$a_0$}&\makebox[3em]{$h$} &\makebox[4em]{$b$ (width)}&\makebox[3em]{$c$}\\[0.3em]\hline
    DCB & 150.0 & 30.5 & 1.50 & 25.0 & -\\[0.3em]
  \hline
    ENF & 101.6 & 35.0 & 2.25 & 25.4 & -\\[0.3em]
    \hline
    MMB & 100.8 & 25.4 & 2.25 & 25.4 & 41.3 \\[0.3em]
    \hline
    \end{tabular}%
  \label{tab:DCBgeo}%
\end{table}%

\begin{table}[h!]
  \centering
  \caption{Material properties for DCB \cite{krueger2015summary}}
    \begin{tabular}{lll}
    \toprule
 \multicolumn{3}{l}{\textbf{T300/1076 Unidirectional graphite/epoxy prepreg}}  \\
 \midrule
    $E_{11}$= 139.4 GPa & $E_{22}$= 10.16 GPa & $E_{33}$ = 10.16 GPa \\[0.3em]
    $\nu_{12}$ = 0.30 & $\nu_{13}$ = 0.30 & $\nu_{23}$ = 0.436  \\[0.3em]
    $G_{12}$ = 4.6 GPa & $G_{13}$ = 4.6 GPa & $G_{23}$ = 3.54 GPa \\[0.5em]
    Fracture toughness data & & \\[0.3em]
    $G_{\rm{Ic}} = 0.170 \, \mathrm{kJ/m^2} $ & $G_{\rm{IIc}} = 0.494 \, \mathrm{kJ/m^2}$ & $\eta = 1.62$ \\ [0.3em]
    Interfacial strength data \cite{turon2010accurate,lu2019cohesive} & & \\[0.3em]
    $\tau_{\rm{Ic}} = 30 \, \mathrm{MPa} $ & $\tau_{\rm{IIc}} = 60 \, \mathrm{MPa}$ &  \\ 
    \bottomrule
    \end{tabular}%
  \label{tab:T300Material}%
\end{table}%

\begin{table}[h!]
  \centering
  \caption{Material properties for ENF and MMB \cite{krueger2015summary}}
    \begin{tabular}{lll}
    \toprule
 \multicolumn{3}{l}{\textbf{IM7/8552 Unidirectional graphite/epoxy prepreg}}  \\
 \midrule
    $E_{11}$= 161 GPa & $E_{22}$= 11.38 GPa & $E_{33}$ = 11.38 GPa \\[0.3em]
    $\nu_{12}$ = 0.32 & $\nu_{13}$ = 0.32 & $\nu_{23}$ = 0.45  \\[0.3em]
    $G_{12}$ = 5.2 GPa & $G_{13}$ = 5.2 GPa & $G_{23}$ = 3.9 GPa \\[0.5em]
    Fracture toughness data & & \\[0.3em]
    $G_{\rm{Ic}} = 0.212 \, \mathrm{kJ/m^2} $ & $G_{\rm{IIc}} = 0.774 \, \mathrm{kJ/m^2}$ & $\eta = 2.1$ \\ [0.3em]
    Interfacial strength data \cite{turon2010accurate,lu2019cohesive} & & \\[0.3em]
    $\tau_{\rm{Ic}} = 30 \, \mathrm{MPa} $ & $\tau_{\rm{IIc}} = 60 \, \mathrm{MPa}$ &  \\ 
    \bottomrule
    \end{tabular}%
  \label{tab:ENFmaterial}%
\end{table}%

\subsubsection{Description of the model}
Since the plies of the benchmarks in this section are all $0\degree$, the symmetric lay-up conditions in equation \ref{eq:MXMYMXY} are met. Therefore, the structural model could simply use one layer of shell elements on each side of the delamination, with one layer of structural CEs in between. The Abaqus solid models are built according to the work of Krueger \cite{krueger2015summary}, except that CEs, instead of VCCT, are used to model delamination.

\subsubsection{Load-displacement curves}

\begin{figure}[h!]
      \centering
	   \begin{subfigure}{0.45\linewidth}
		\includegraphics[width=\linewidth]{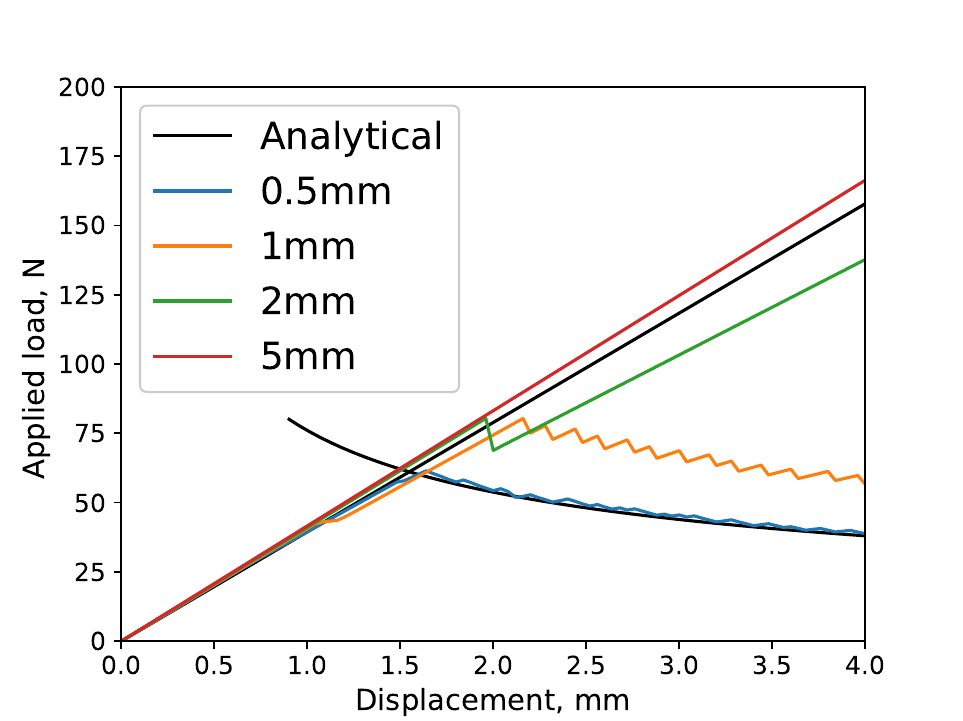}
		\caption{DCB-Solid}
		\label{fig:dcb-solid}
	   \end{subfigure}
	   \begin{subfigure}{0.45\linewidth}
		\includegraphics[width=\linewidth]{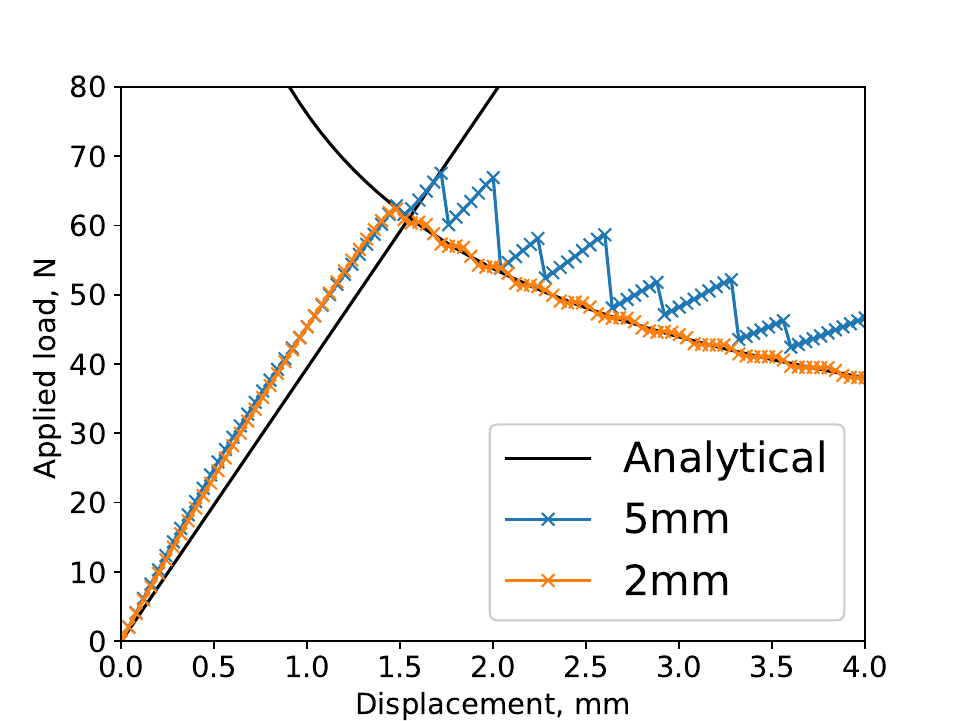}
		\caption{DCB-Structural}
		\label{fig:dcb-structural}
	    \end{subfigure}
	\vfill
	     \begin{subfigure}{0.45\linewidth}
		 \includegraphics[width=\linewidth]{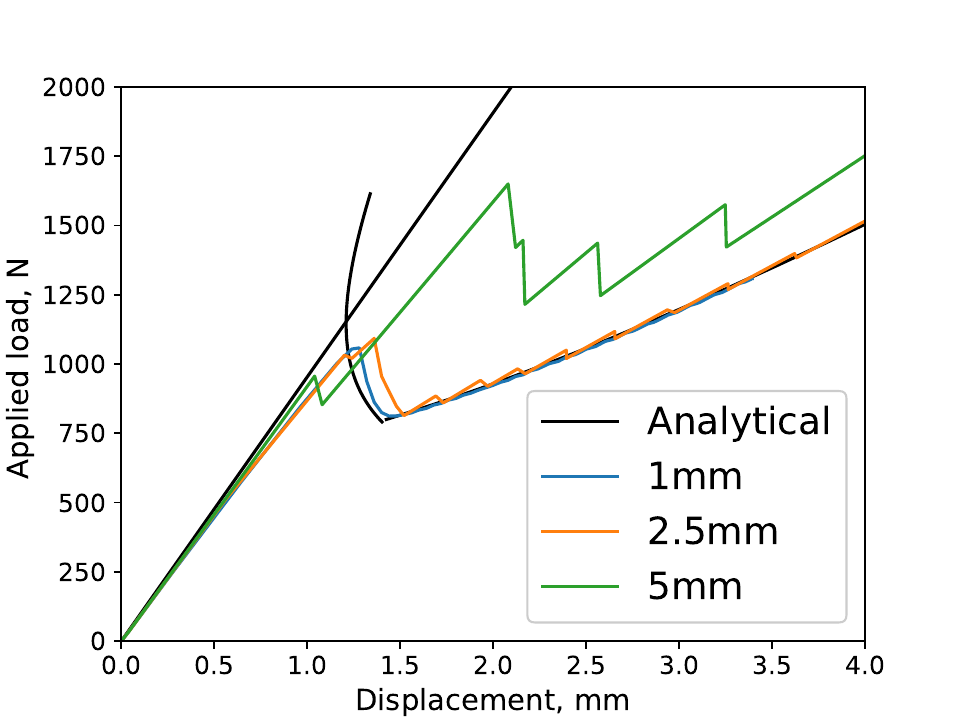}
		 \caption{ENF-Solid}
		 \label{fig:enf-solid}
	      \end{subfigure}
	       \begin{subfigure}{0.45\linewidth}
		  \includegraphics[width=\linewidth]{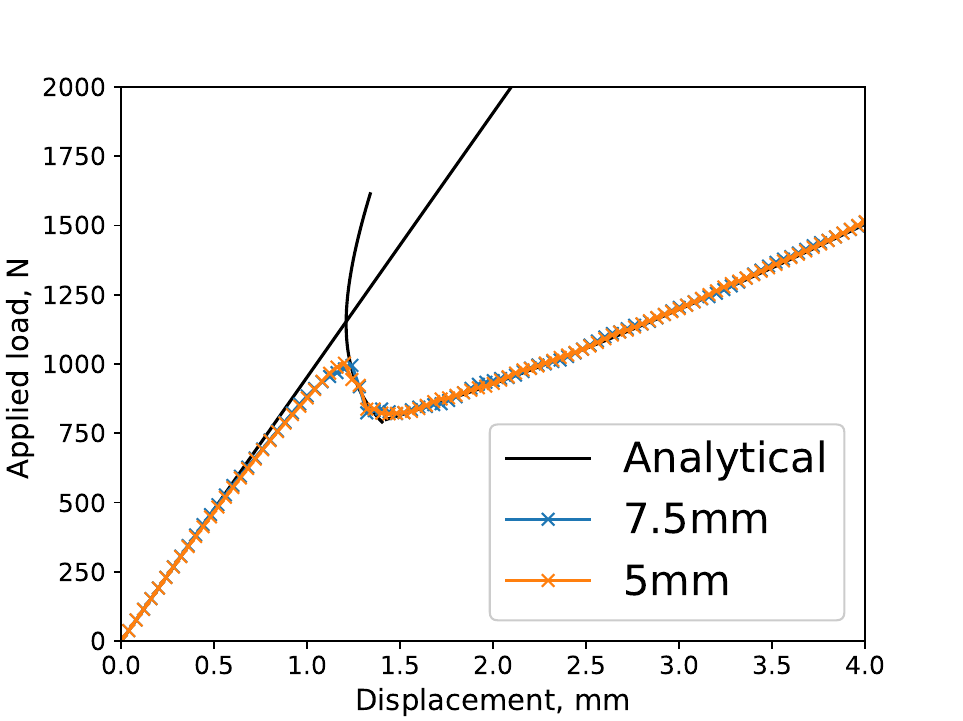}
		  \caption{ENF-Structural}
		  \label{fig:enf-structural}
	       \end{subfigure}
      \vfill  
      	     \begin{subfigure}{0.45\linewidth}
		 \includegraphics[width=\linewidth]{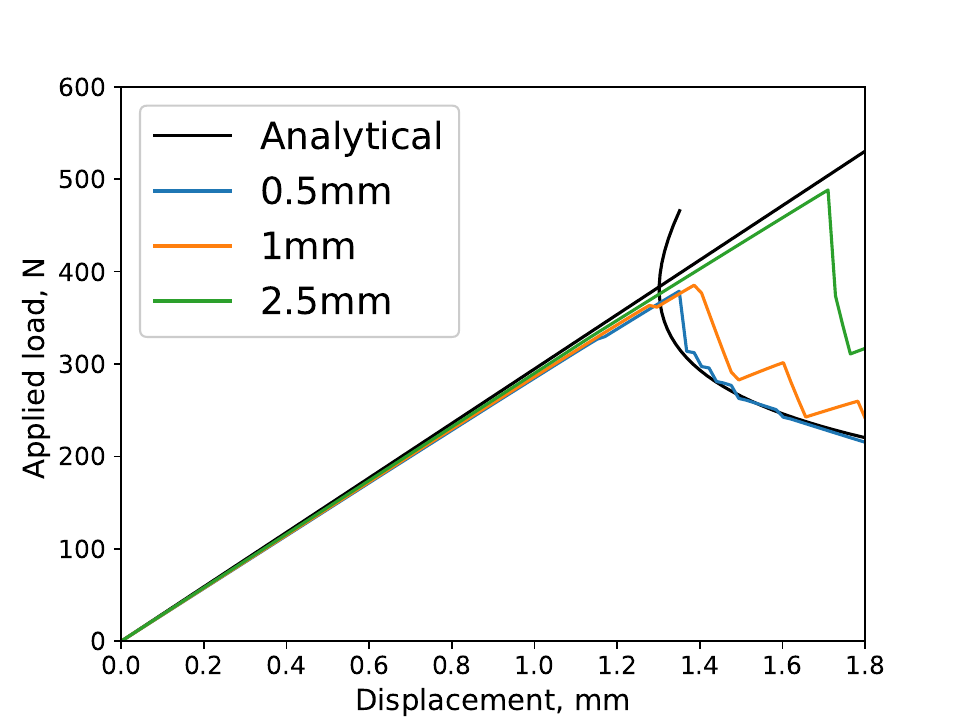}
		 \caption{MMB-Solid}
		 \label{fig:mmb-solid}
	      \end{subfigure}
	       \begin{subfigure}{0.45\linewidth}
		  \includegraphics[width=\linewidth]{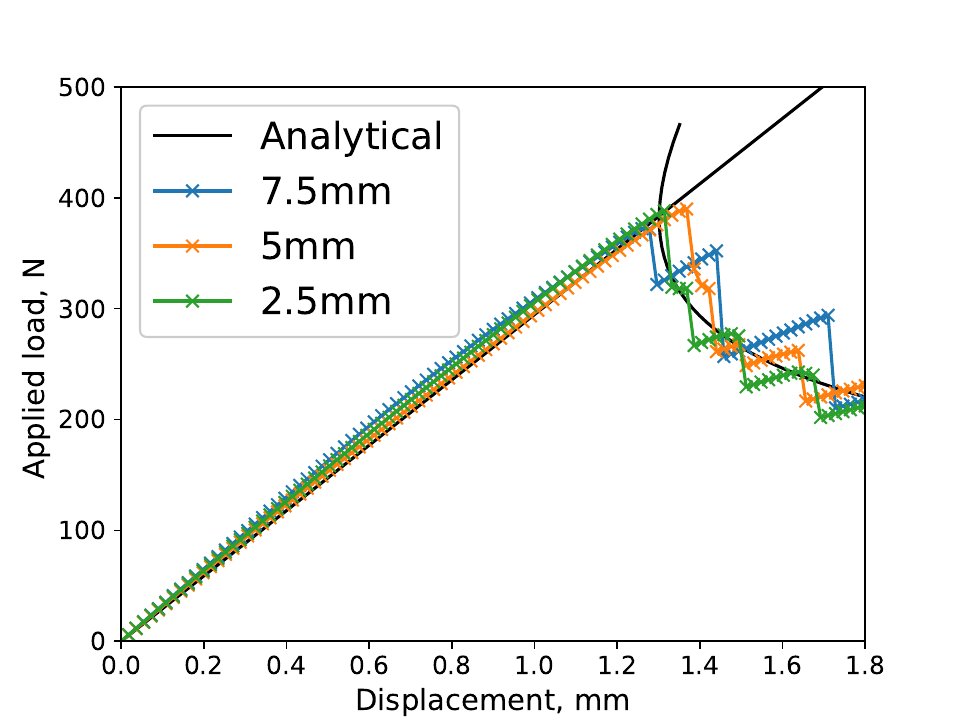}
		  \caption{MMB-Structural}
		  \label{fig:mmb-structural}
	       \end{subfigure}
	\caption{Results of the DCB, ENF, and MMB tests}
	\label{fig:result_UDbenchmarks}
\end{figure}
The load-displacement curves obtained from the simulations are shown in Figure \ref{fig:result_UDbenchmarks}. The designations ``-solid" and ``-structural" denote the results of the traditional solid element model and those of the proposed structural CE model, respectively. The results on meshes of different element sizes are plotted together with the analytical solutions \cite{williams1989fracture,ASTMD5528,ASTMD6671/D6671M}.

The DCB results in Figure \ref{fig:result_UDbenchmarks} indicated that the standard model with solid elements requires that the element size should not exceed 0.5 mm. When the element sizes are greater than or equal to 1 mm, the simulation results differ greatly from the analytical solution. The error on peak load exceeds 30\%, and the post-peak curve also stays way above the analytical one. With the proposed structural model, the results on 2-mm mesh remain in close agreement with the analytical solution, with a less than 3\% error on the peak load. The post-peak curve also closely follows the analytical curve. The slight over-prediction of the stiffness is expected as a result of neglecting transverse shear in the Kirchhoff-Love shell elements. The analytical solution based on the corrected beam theory, however, includes such transverse shear effect \cite{williams1989fracture}. Even on a 5-mm mesh, the structural model can predict the peak load fairly well despite the post-peak oscillations right above the analytical curve. Such oscillations are due to the larger spacing between the integration points on coarser meshes. 

In the ENF case, the solid element model on 1-mm mesh could capture the peak load correctly, thanks to the larger cohesive zone in Mode II delamination than in Mode I. However, the result on 2.5-mm mesh already shows a clear drift towards over-prediction. As the element size increases to 5 mm, the peak load and post-peak response again become severely over-predicted. Correspondingly, if the structural model is used on the 5-mm mesh, the predicted curve remains close to the analytical solution throughout the loading history. Even the 7.5-mm structural model manages to capture the load-displacement response very accurately. The slight under-prediction of the peak load is expected, as the analytical curve is based on Linear Elastic Fracture Mechanics, which ignores the presence of material softening (i.e. the cohesive zone) at the crack tip.  

In the MMB case, very similar trends can be observed. The solid model on the 2.5-mm mesh cannot capture the correct load-displacement response, while the structural model's predictions on the 2.5-mm and 5-mm meshes oscillates closely around the analytical curve. Even the 7.5-mm structural model predicts the peak load correctly, albeit with bigger oscillations during the load drop section due to the coarser distribution of integration points in larger elements. 

 


\subsubsection{Computational performances}

\begin{table}[h]
	\centering
	\caption{Comparison of CPU time (unit: second)}
	\begin{tabular}{l l l l}
		\toprule
         & DCB  & ENF & MMB \\[0.3em]
         \midrule
         Solid model (mesh size) & 5311.8 (0.5mm) & 4239.7 (1mm) & 5560.7 (0.5mm)\\[0.3em]
         \midrule
         Structural model (mesh size) & 197.15 (5mm) & 285.33 (7.5mm)& 370.02 (7.5mm)\\[0.3em]
          & 836.76 (2mm) & 736.41 (5mm)& 474.79 (5mm)\\[0.3em]
         \midrule
         Reduction by structural & 96.3\% & 93.3\% & 93.3\% \\[0.3em]
          & 84.2\% & 82.6\% & 91.5\% \\[0.3em]
         \bottomrule 
	\end{tabular}
	\label{tab:Computational performances}
\end{table}

By allowing larger elements to be used, the proposed structural model is able to reduce the computational time of the delamination simulations considerably. The results of the structural model are compared against those of the solid element model. The comparison on CPU time is reported in Table \ref{tab:Computational performances}. It can be seen that the structural model can reduce the computational time in all three problems by more than 90\%, while retaining accurate predictions of the peak loads and the overall load-drop curves.

\subsection{Multi-directional laminate benchmark: SLB}
\subsubsection{Description of the SLB test}
\begin{figure}[h]
	\centering
	\includegraphics[width=0.9\linewidth]{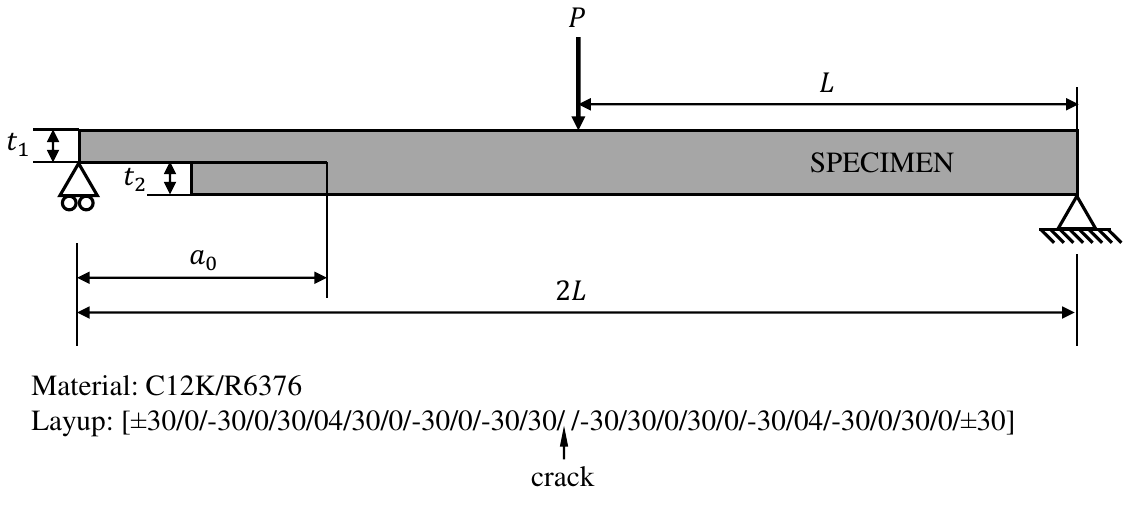}
	\caption{SLB specimen \cite{davidson1995three}}
	\label{fig:SLB}
\end{figure}

The SLB specimen is shown in Figure~\ref{fig:SLB}, with its geometrical parameters specified in Table~\ref{tab:SLBgeo}. Unlike in the previous benchmarks, the ply angles here are no longer all $0\degree$. The material of the SLB model is C12K/R6376 and its properties are shown in Table~\ref{tab:C12K/R6376Material}. The value of $\tau_{\rm{IIc}}$ cannot be found in the literature. According to the review in Ref.~\cite{lu2019cohesive}, the ratio of $\tau_{\rm{IIc}}$ to $\tau_{\rm{Ic}}$ ranges from 1.25 to 2. Hence, the value of $\tau_{\rm{IIc}}$ here is estimated to be 38 MPa, 1.5 times that of $\tau_{\rm{Ic}}$.

\begin{table}[h]
  \centering
  \caption{Geometric parameters for the SLB specimen}
    \begin{tabular}{llllll}
    \toprule
    Parameter & $2L$ & $a_0$ & $t_1$ & $t_2$ & $b$ (width)  \\[0.3em]
 \midrule
    value (mm) & 177.8 & 60 & 2 & 2 & 25.4 \\[0.3em]
    \bottomrule
    \end{tabular}%
  \label{tab:SLBgeo}%
\end{table}

\begin{table}[h]
  \centering
  \caption{C12K/R6376 material properties for SLB specimen \cite{krueger2015summary}}
    \begin{tabular}{lll}
    \toprule
    $E_{11}$= 146.9 GPa & $E_{22}$= 10.6 GPa & $E_{33}$ = 10.6 GPa \\[0.3em]
    $\nu_{12}$ = 0.33 & $\nu_{13}$ = 0.33 & $\nu_{23}$ = 0.33  \\[0.3em]
    $G_{12}$ = 5.45 GPa & $G_{13}$ = 5.45 GPa & $G_{23}$ = 3.99 GPa \\[0.5em]
    Fracture toughness data & & \\[0.3em]
    $G_{\rm{Ic}} = 0.34 \, \mathrm{kJ/m^2} $ & $G_{\rm{IIc}} = 1.286 \, \mathrm{kJ/m^2}$ & $\eta = 3.39$ \\ [0.3em]
    Interfacial strength data & & \\[0.3em]
    $\tau_{\rm{Ic}} = 25 \, \mathrm{MPa} $ \cite{bruyneel2009modeling} & $\tau_{\rm{IIc}} = 38 \, \mathrm{MPa}${*} &  \\ 
    \bottomrule
    & & {*} Estimated \cite{lu2019cohesive} \\
    \end{tabular}%
  \label{tab:C12K/R6376Material}%
\end{table}%

\subsubsection{Description of the model}

\begin{figure}[h!]
	\centering
	\includegraphics[width=0.9\linewidth]{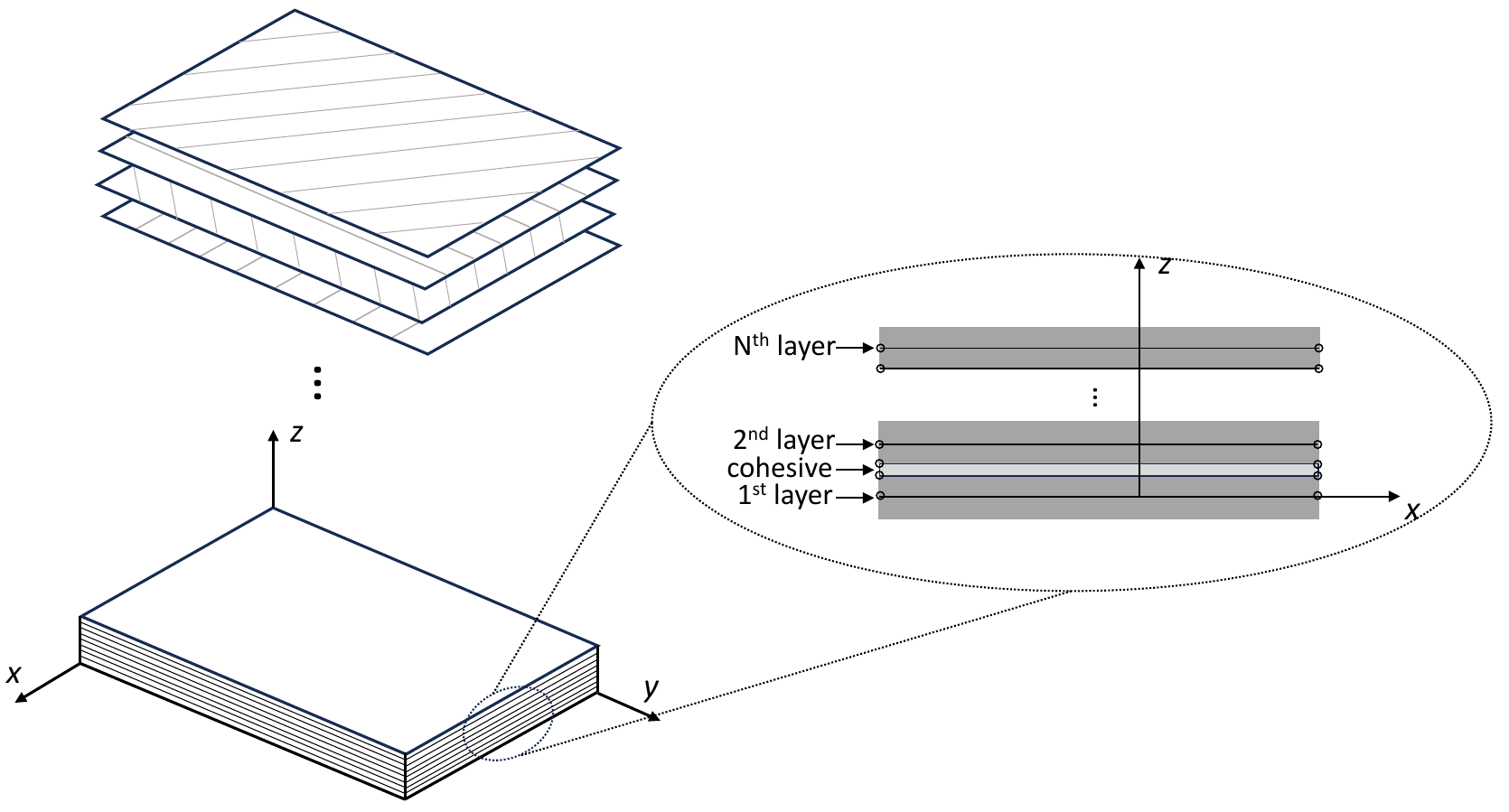}
	\caption{Layer-wise model for SLB}
    \label{fig: SLB model building}
\end{figure}

Since the SLB specimen is a multi-directional laminate and its lay-up is not symmetric, we can no longer model the entire laminate with a single layer of shell elements on each side of the delamination, as done in the unidirectional models. Therefore, the SLB model uses one layer of shell elements for each ply and one layer of structural CEs between every two plies (c.f., Figure~\ref{fig: SLB model building}). 

\subsubsection{Load-displacement curves}

\begin{figure}[h!]
      \centering
	   \begin{subfigure}{0.45\linewidth}
		\includegraphics[width=\linewidth]{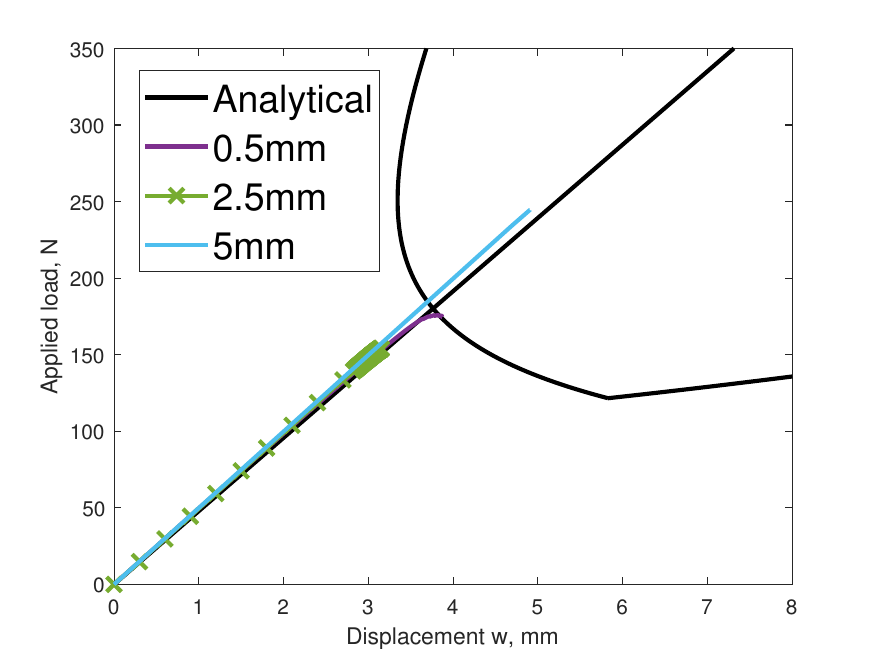}
		\caption{SLB-Solid}
		\label{fig:slb-solid}
	   \end{subfigure}
	   \begin{subfigure}{0.45\linewidth}
		\includegraphics[width=\linewidth]{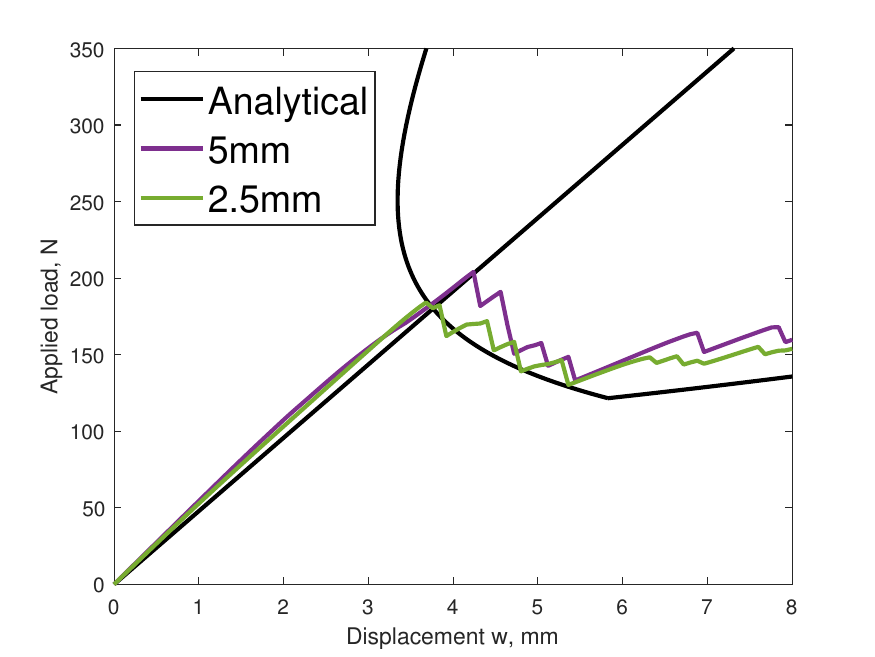}
		\caption{SLB-Structural}
		\label{fig:slb-structural}
	    \end{subfigure}
	\caption{Load-displacement predictions of the SLB}
	\label{fig:SLB_curve}
\end{figure}

In Figure \ref{fig:SLB_curve}, the load-displacement curves are reported for the solid and structural models. Both are compared with the analytical solution derived in \ref{sec:analytical-benchmarks}. All the load-displacement predictions by the solid model could not reach numerical convergence within the allowable settings on solver iterations. As a result, the load-decreasing section of the curve could not be obtained. From the left figure, it can be seen that only the 0.5-mm solid model can capture the correct peak load. For the 2.5-mm solid model, the peak load is much lower than the analytical solution due to premature divergence of analysis. However, the peak load of the model on the 5-mm mesh is much larger than the analytical solution, in line with the solid model's performance in the unidirectional benchmarks. The right figure shows that all the structural models can obtain converged solutions throughout the full loading history. Overall speaking, the predicted curves of the structural models follow well the analytical solution. On the 2.5-mm mesh, the error of peak load is 2\%, the same level of accuracy as that of the unidirectional structural model. On the 5-mm mesh, the predicted peak load is larger than that of the analytical solution by 12.5\%. However, the post-peak part of the curve quickly converges to that of the 2.5-mm mesh result. It should be noted that the initial stiffness of the structural models is 5\% larger than that of the analytical solution. The reason is that the compliance $C$ obtained by equation \ref{eq: compliance of SLB} is calculated by simplifying the SLB model into a 2D beam model (c.f., Figure~\ref{fig: SLB_analytical}), where the neutral axis is assumed to be aligned with the centerline of the uncracked region for simplicity. However, the actual neutral axis of the SLB model should be slightly above the assumed one, as only the top part of the cracked region would be under bending. This discrepancy would cause the analytical stiffness to be slightly lower than that in the actual situation. Another discrepancy to note is that when the crack length $a$ is longer than $L$, the load increases again with increasing displacement (c.f., \ref{sec:analytical-benchmarks}, curve DFE). For this part, the predicted curves of the structural models are above the analytical curve. This discrepancy is however expected. As the crack length $a$ increases with delamination propagation, the mixed-mode ratio $B$ is found to also increase, particularly when $a>L$ \cite{krueger2020benchmark}, which causes the critical energy release rate $G_c$ to increase as well. However, the analytical solution in \ref{sec:analytical-benchmarks} assumes a fixed mixed-ratio, which underestimates $G_c$ for $a>L$, hence resulting in the analytical curve being below the numerical ones. The analytical curve could be corrected, should an analytical expression of $G_c$ in terms of $a$ becomes available. 



In addition, it needs to be emphasized that the difficulty of convergence leads to a significant increase in computational time. Due to this reason, artificial viscosity would be needed to continue the simulation through the convergence difficulties. However, this would introduce a fitting parameter which generally requires trial and error to properly set its value. This suggests that the proposed structural model not only allows coarser meshes with faster computations but also ensures better numerical stability than the solid element model.


\section{Summary and conclusions}
\label{sec:conclusions}
This work aims to establish a state-of-the-art numerical method to simulate the delamination of composite laminates aimed at overcoming the cohesive zone limit on mesh density. The Kirchhoff-Love triangular cubic plate element from Allman \cite{allman1976simple} has been extended for the modelling of symmetric laminate shells. A structural CE, conforming with the shell element, has been developed to model delamination between the shells. The combination of the Kirchhoff-Love shell element and the structural CE is shown to be a powerful new method that overcomes the cohesive zone limit and models delamination with high accuracy and efficiency under different loading conditions. This capability is expected to make a strong impact in the composites modeling community, as the cohesive zone limit has been a long-lasting and well-known problem for delamination modeling. 

This method has also been validated on the DCB, ENF, MMB, and SLB problems. The results show excellent agreement with the analytical solutions. By comparing the results with those from the solid element model, the structural model has shown the following advantages: 1) it can accurately predict the load-displacement curves with significantly coarser meshes than the solid element model, allowing 5-mm elements to be used instead of the 0.5-mm elements in the latter; 2) its computational speed is much faster than that of the solid element model, achieving more than 90\% reduction in CPU time; and 3) it has better numerical stability than the solid element model, reaching convergence without needing artificial viscosity for stabilization.

Future work includes the incorporation of intralaminar damage in the structural model. Since Kirchhoff-Love thin shell elements are used to model the plies, the transverse strains are ignored. This will impact the prediction of intralaminar damage in the model because an accurate three-dimensional damage criterion would require out-of-plane stresses and/or strains as inputs. However, since the layer-by-layer modelling method used in this work employs CEs between every two plies, the required out-of-plane stresses can be obtained by extracting the tractions of the CEs and interpolating them between the layers \cite{russo2020overcoming}. This is the subject of an ongoing work in the group.

\section*{Acknowledgements}
The authors would like to acknowledge the useful discussions with Giorgio Tosti Balducci, PhD candidate in the same group. The first author would like to acknowledge funding support from the China Scholarship Council (No.201906290034) for this research.

\bibliographystyle{elsarticle-num-names} 
\bibliography{cas-refs}
\appendix

\section{$\bH$ matrix}
\label{sec:H-matrix}
Before we calculate $U_0$ in equation \ref{eq:U0}, we need to calculate six parts. The first part is:
\begin{align}
    \frac{D_{11}}{2} \left(\frac{\partial^2 w}{\partial x^2} \right)^2 = D_{11} (2\alpha_1^2 + 12x\alpha_1\alpha_4 + 4y\alpha_1\alpha_5
     + 18x^2\alpha_4^2 + 12xy\alpha_4\alpha_5 + 2y^2\alpha_5^2)
\end{align}

Based on equation \ref{eq:U0=0.5aHa}, we can obtain:
\begin{align}\label{eq:appendix_H1}
    \iint_{A} \frac{D_{11}}{2} \left(\frac{\partial^2 w}{\partial x^2} \right)^2 \, \dd x \dd y = \frac{1}{2} \balpha \transpose \, \bH_1 \, \balpha
\end{align}
where $\bH_1$ is:
\begin{align}
    \bH_1 = \iint_{A} D_{11} 
    \begin{bmatrix}
    4 & 0 & 0 & 12x   & 4y   & 0 & 0 \\[0.5em]
      & 0 & 0 & 0     & 0    & 0 & 0 \\[0.5em]
      &   & 0 & 0     & 0    & 0 & 0 \\[0.5em]
      &   &   & 36x^2 & 12xy & 0 & 0 \\[0.5em]
      & \rm{Symmetric}  &   &       & 4y^2 & 0 & 0 \\[0.5em]
      &   &   &       &      & 0 & 0 \\[0.5em]
      &   &   &       &      &   & 0 
    \end{bmatrix} \dd x \dd y
\end{align}

The second part is:
\begin{align}
    \frac{D_{22}}{2}\left(\frac{\partial^2 w}{\partial y^2} \right)^2 = D_{22} (2\alpha_3^2 + 4x\alpha_3\alpha_6 + 12y\alpha_3\alpha_7 
    + 2x^2\alpha_6^2 + 12xy\alpha_6\alpha_7 + 18y^2\alpha_7^2)
\end{align}

Then, an equation which is similar as equation \ref{eq:appendix_H1} in terms of $\bH_2$ can be obtained:
\begin{align}
    \iint_{A} \frac{D_{22}}{2} \left(\frac{\partial^2 w}{\partial x^2} \right)^2 \, \dd x \dd y = \frac{1}{2} \balpha \transpose \, \bH_2 \, \balpha
\end{align}
where $\bH_2$ is: 
\begin{align}
    \bH_2 = \iint_{A} D_{22} 
    \begin{bmatrix}
    0 & 0 & 0 & 0   & 0    & 0    & 0    \\[0.5em]
      & 0 & 0 & 0   & 0    & 0    & 0    \\[0.5em]
      &   & 4 & 0   & 0    & 4x   & 12y  \\[0.5em]
      &   &   & 0   & 0    & 0    & 0    \\[0.5em] 
      &\rm{Symmetric}   &   &     & 0    & 0    & 0    \\[0.5em] 
      &   &   &     &      & 4x^2 & 12xy \\[0.5em]
      &   &   &     &      & 0    & 36y^2 
    \end{bmatrix} \dd x \dd y
\end{align}

The third part is:
\begin{align}
    D_{12}\frac{\partial^2 w}{\partial x^2}\frac{\partial^2 w}{\partial y^2} =  &D_{12}(4\alpha_1\alpha_3 + 4x\alpha_1\alpha_6 + 12x\alpha_3\alpha_4 + 12y\alpha_1\alpha_7 + 4y\alpha_3\alpha_5 \nonumber\\[0.5em]
    &+ 12x^2\alpha_4\alpha_6 + 12y^2\alpha_5\alpha_7 + 36xy\alpha_4\alpha_7 + 4xy\alpha_5\alpha_6)
\end{align}
and $\bH_3$ can be obtained by using the same method:
\begin{align}
    \bH_3 = \iint_{A} D_{12} 
    \begin{bmatrix}
    0 & 0 & 4 & 0   & 0    & 4x    & 12y    \\[0.5em]
      & 0 & 0 & 0   & 0    & 0     & 0      \\[0.5em]
      &   & 0 & 12x & 4y   & 0     & 0      \\[0.5em]  
      &   &   & 0   & 0    & 12x^2 & 36xy   \\[0.5em]
      &\rm{Symmetric}   &  &       & 0    & 4xy    & 12y^2    \\[0.5em]
      &   &   &     &      & 0     & 0      \\[0.5em]
      &   &   &     &      &       & 0 
    \end{bmatrix} \dd x \dd y
\end{align}

Therefore, the $\bH$ matrix of $4^\mathrm{th}$, $5^\mathrm{th}$ and $6^\mathrm{th}$ part are expressed by the following expressions:
\begin{align}
    \bH_4 = \iint_{A} D_{16} 
    \begin{bmatrix} 
    0 & 4 & 0 & 0   & 8x    & 8y    & 0    \\[0.5em]
      & 0 & 0 & 12x & 4y    & 0     & 0    \\[0.5em]
      &   & 0 & 0   & 0     & 0     & 0    \\[0.5em] 
      &   &   & 0   & 24x^2 & 24xy  & 0    \\[0.5em]   
      &\rm{Symmetric}   &   &     & 16xy  & 8y^2  & 0    \\[0.5em]
      &   &   &     &       & 0     & 0    \\[0.5em]
      &   &   &     &       &       & 0   
    \end{bmatrix} \dd x \dd y
\end{align}

\begin{align}
    \bH_5 = \iint_{A} D_{26}
    \begin{bmatrix}
    0 & 0 & 0 & 0   & 0     & 0     & 0     \\[0.5em]
      & 0 & 4 & 0   & 0     & 4x    & 12y   \\[0.5em]
      &   & 0 & 0   & 8x    & 8y    & 0     \\[0.5em]
      &   &   & 0   & 0     & 0     & 0     \\[0.5em]
      &\rm{Symmetric}   &   &     & 0     & 8x^2  & 24xy  \\[0.5em]
      &   &   &     &       & 16xy   & 24y^2 \\[0.5em]
      &   &   &     &       &       & 0  
    \end{bmatrix}\dd x \dd y
\end{align}
and
\begin{align}
    \bH_6 = \iint_{A} D_{66} 
    \begin{bmatrix} 
    0 & 0 & 0 & 0   & 0     & 0     & 0    \\[0.5em]
      & 4 & 0 & 0   & 8x    & 8y    & 0    \\[0.5em]
      &   & 0 & 0   & 0     & 0     & 0    \\[0.5em] 
      &   &   & 0   & 0     & 0     & 0    \\[0.5em]   
      &\rm{Symmetric}   &   &     & 16x^2  & 16xy  & 0    \\[0.5em]
      &   &   &     &       & 16y^2 & 0    \\[0.5em]
      &   &   &     &       &       & 0   
    \end{bmatrix} \dd x \dd y
\end{align}

The $\bH$ matrix can be written as the sum of $\bH_i$ $(i=1,2,...,6)$:
\begin{align}
    \bH = \bH_1 + \bH_2 + \bH_3 + \bH_4 + \bH_5+ \bH_6
\end{align}
where the integral for each item of the $\bH$ matrix can be calculated by using the formulas given in Reference~\cite{bell1969refined}.

\section{$\bB$ matrix}
\label{sec:B-matrix}
Plugging the expression of $w$ (equation~\ref{eq:cubicw}) into equation \ref{eq:MXMYMXY}, we can express the moment resultants as:
\begin{align}
    M_x = &-2D_{11}\alpha_1 -2D_{16}\alpha_2 -2D_{12}\alpha_3 - 6xD_{11}\alpha_4 - (2yD_{11}+4xD_{16})\alpha_5  \nonumber\\[0.5em]
    &-(2xD_{12}+4yD_{16})\alpha_6 - 6yD_{12}\alpha_7
\end{align}
which in matrix form is:
\begin{align}\label{eq:Appendix-B-Mx}
       M_x = \bB_{M_x}\transpose \, \balpha
\end{align}
where $\bB_{M_x}\transpose$ is:
\begin{align}\label{eq:appendixB_Bmx}
    \bB_{M_x}\transpose = [ &-2D_{11},\, -2D_{16}, \, -2D_{12} ,\,-6xD_{11}, \,-(2yD_{11}+4xD_{16}), \nonumber \\[0.5em]
    & -(2xD_{12}+4yD_{16}),  \,-6yD_{12} ] 
\end{align}

Similarly, for $M_y$, we have:
\begin{align}
    M_y = \bB_{M_y}\transpose \, \balpha
\end{align}
where $\bB_{M_y}\transpose$ is:
\begin{align}\label{eq:appendixB_Bmy}
    \bB_{M_y}\transpose = 
    [&-2D_{12},\,  -2D_{26},\, -2D_{22},\, -6xD_{12},\, - (2yD_{12}+4xD_{26}), \nonumber \\[0.5em]
    & - (2xD_{22}+4yD_{26}),\,-6yD_{22} ] 
\end{align}

For $M_{xy}$, we can obtain:
\begin{align}
    M_{xy} = \bB_{M_{xy}}\transpose \, \balpha
\end{align}
where the matrix $\bB_{M_{xy}}\transpose$ is:
\begin{align}
    \bB_{M_{xy}}\transpose = [
    &-2D_{16},\, -2D_{66} ,\,  -2D_{26},\,  - 6xD_{16} ,\, - (4xD_{66}+2yD_{16}), \nonumber \\[0.5em]
    & - (4yD_{66}+2xD_{26}),\,  -6yD_{26}]
\end{align}

From equation~\ref{eq:MXMYMXY}, we can calculate the derivative of the moment:
\begin{align}
    \frac{\partial M_x}{\partial x} &= - D_{11} \frac{\partial^3 w}{\partial x^3}  - D_{12} \frac{\partial^3 w}{\partial x \partial y^2} - 2D_{16}\frac{\partial^3 w}{\partial x^2\partial y} \nonumber\\[0.5em]
    \frac{\partial M_x}{\partial y} &= - D_{11} \frac{\partial^3 w}{\partial x^2\partial y}  - D_{12} \frac{\partial^3 w}{\partial y^3} - 2D_{16}\frac{\partial^3 w}{\partial x \partial y^2}\nonumber  \\[0.5em]    
    \frac{\partial M_y}{\partial x} &= - D_{12} \frac{\partial^3 w}{\partial x^3}  - D_{22} \frac{\partial^3 w}{\partial x \partial y^2}  - 2D_{26}\frac{\partial^3 w}{\partial x^2\partial y}\nonumber\\[0.5em]
    \frac{\partial M_y}{\partial y} &= - D_{12} \frac{\partial^3 w}{\partial x^2\partial y}  - D_{22} \frac{\partial^3 w}{\partial y^3} - 2D_{26}\frac{\partial^3 w}{\partial x \partial y^2} \nonumber\\[0.5em]
    \frac{\partial M_{xy}}{\partial x} &= - D_{16} \frac{\partial^3 w}{\partial x^3}  - D_{26} \frac{\partial^3 w}{\partial x \partial y^2} - 2D_{66} \frac{\partial^3 w}{\partial x^2 \partial y} \nonumber\\[0.5em]
    \frac{\partial M_{xy}}{\partial y} &= - D_{16} \frac{\partial^3 w}{\partial x^2\partial y}  - D_{26} \frac{\partial^3 w}{\partial y^3}- 2D_{66} \frac{\partial^3 w}{\partial x \partial y^2} 
\end{align}

Using equation \ref{eq:moment1-1}, we calculate $M_n$ for node $1$ and node $2$ on the side 1-2 as: 
\begin{align}\label{eq:Mn_12}
    M_n^{12} = \cos^2{\gamma_{12}}\bB_{M_x^1}\transpose\balpha + \sin^2{\gamma_{12}}\bB_{M_y^1}\transpose\balpha + \sin{2\gamma_{12}}\bB_{M_{xy}^1}\transpose\balpha\\[0.5em]
    M_n^{21} = \cos^2{\gamma_{12}}\bB_{M_x^2}\transpose\balpha + \sin^2{\gamma_{12}}\bB_{M_y^2}\transpose\balpha + \sin{2\gamma_{12}}\bB_{M_{xy}^2}\transpose\balpha    
\end{align}
where $\bB_{M_x^j}\transpose$ means equation \ref{eq:appendixB_Bmx} evaluated at the coordinates of node $j$ and $\bB_{M_y^j}\transpose$ means equation \ref{eq:appendixB_Bmy} evaluated at the coordinates of node $j$, respectively. And the corresponding $\bB$ matrix is:
\begin{align}
    \bB_{M_n^{12}}\transpose = \cos^2{\gamma_{12}}\bB_{M_x^1}\transpose + \sin^2{\gamma_{12}}\bB_{M_y^1}\transpose + \sin{2\gamma_{12}}\bB_{M_{xy}^1}\transpose \\[0.5em]
    \bB_{M_n^{21}}\transpose = \cos^2{\gamma_{12}}\bB_{M_x^2}\transpose + \sin^2{\gamma_{12}}\bB_{M_y^2}\transpose + \sin{2\gamma_{12}}\bB_{M_{xy}^2}\transpose 
\end{align}

Similarly, $M_n$ for the side 2-3 is: 
\begin{align}\label{eq:Mn_23}
    M_n^{23} = \cos^2{\gamma_{23}}\bB_{M_x^2}\transpose\balpha +  \sin^2{\gamma_{23}}\bB_{M_y^2}\transpose\balpha +  \sin{2\gamma_{23}}\bB_{M_{xy}^2}\transpose\balpha\\[0.5em]
    M_n^{32} = \cos^2{\gamma_{23}}\bB_{M_x^3}\transpose\balpha +  \sin^2{\gamma_{23}}\bB_{M_y^3}\transpose\balpha +  \sin{2\gamma_{23}}\bB_{M_{xy}^3}\transpose\balpha  
\end{align}
And the corresponding $\bB$ matrix is:
\begin{align}
    \bB_{M_n^{23}}\transpose = \cos^2{\gamma_{23}}\bB_{M_x^2}\transpose + \sin^2{\gamma_{23}}\bB_{M_y^2}\transpose + \sin{2\gamma_{23}}\bB_{M_{xy}^2}\transpose \\[0.5em]
    \bB_{M_n^{32}}\transpose = \cos^2{\gamma_{23}}\bB_{M_x^3}\transpose + \sin^2{\gamma_{23}}\bB_{M_y^3}\transpose + \sin{2\gamma_{23}}\bB_{M_{xy}^3}\transpose 
\end{align}

Finally, $M_n$ for the side 3-1 is: 
\begin{align}\label{eq:Mn_31}
    M_n^{31} = \cos^2{\gamma_{31}}\bB_{M_x^3}\transpose\balpha +  \sin^2{\gamma_{31}}\bB_{M_y^3}\transpose\balpha +  \sin{2\gamma_{31}}\bB_{M_{xy}^3}\transpose\balpha\\[0.5em]
    M_n^{13} = \cos^2{\gamma_{31}}\bB_{M_x^1}\transpose\balpha +  \sin^2{\gamma_{31}}\bB_{M_y^1}\transpose\balpha +  \sin{2\gamma_{31}}\bB_{M_{xy}^1}\transpose\balpha 
\end{align}
And the corresponding $\bB$ matrix is:
\begin{align}
    \bB_{M_n^{31}}\transpose = \cos^2{\gamma_{31}}\bB_{M_x^3}\transpose + \sin^2{\gamma_{31}}\bB_{M_y^3}\transpose + \sin{2\gamma_{31}}\bB_{M_{xy}^3}\transpose \\[0.5em]
    \bB_{M_n^{13}}\transpose = \cos^2{\gamma_{31}}\bB_{M_x^1}\transpose + \sin^2{\gamma_{31}}\bB_{M_y^1}\transpose + \sin{2\gamma_{31}}\bB_{M_{xy}^1}\transpose 
\end{align}

Before calculating $R_N$, we need to calculate $M_{ns}$ at first as shown in equation \ref{eq:R1,R2,R3}. For node 1:
\begin{align}
    M_{ns}^{12} = \left[\frac{1}{2}\left(\bB_{M_y^1}\transpose -\bB_{M_x^1}\transpose\right)\sin 2\gamma_{12} + \bB_{M_{xy}^1}\transpose  \cos 2\gamma_{12}\right]\,\balpha\\[0.5em]
    M_{ns}^{13} = \left[\frac{1}{2}\left(\bB_{M_y^1}\transpose -\bB_{M_x^1}\transpose\right)\sin 2\gamma_{31} +\bB_{M_{xy}^1}\transpose  \cos 2\gamma_{31} \right]\,\balpha
\end{align}

Thus, plugging the above two terms into equation \ref{eq:R1,R2,R3}, $R_1$ is:
\begin{align}
    R_1 &= \left[\frac{1}{2}(\sin 2\gamma_{12} - \sin 2\gamma_{31})\left(\bB_{M_y^1}\transpose -\bB_{M_x^1}\transpose\right)\nonumber + (\cos 2\gamma_{12}-\cos 2\gamma_{31})\bB_{M_{xy}^1}\transpose \right]\,\balpha \nonumber\\[0.5em]
    &= \bB_{R_1} \, \balpha
\end{align}
where the matrix $\bB_{R_1}$ is:
\begin{align}\label{eq:BR1}
    \bB_{R_1}\transpose  = \frac{1}{2}\left(\sin 2\gamma_{12} - \sin 2\gamma_{31}\right)\left(\bB_{M_y^1}\transpose -\bB_{M_x^1}\transpose\right)+ (\cos 2\gamma_{12}-\cos 2\gamma_{31})\bB_{M_{xy}^1}\transpose
\end{align}

Similarly, the matrix $\bB_{R_2}$ related to the node 2:
\begin{align}\label{eq:BR2}
    \bB_{R_2}\transpose = \frac{1}{2}\left(\sin 2\gamma_{23} - \sin 2\gamma_{12}\right)\left(\bB_{M_y^2}\transpose -\bB_{M_x^2}\transpose\right) + (\cos 2\gamma_{23}-\cos 2\gamma_{12})\bB_{M_{xy}^2}\transpose        
\end{align}

At last, the matrix $\bB_{R_3}$:
\begin{align}\label{eq:BR3}
    \bB_{R_3}\transpose = \frac{1}{2}\left(\sin 2\gamma_{31} - \sin 2\gamma_{23}\right)\left(\bB_{M_y^3}\transpose -\bB_{M_x^3}\transpose\right) + (\cos 2\gamma_{31}-\cos 2\gamma_{23})\bB_{M_{xy}^3}\transpose        
\end{align}

In order to simplify the writing of the above formula, we define the following notations:
\begin{align}
    \overline{S}_1 = \sin{2\gamma_{12}} - \sin{2\gamma_{31}}, \quad \overline{C}_1 = \cos{2\gamma_{12}} - \cos{2\gamma_{31}} \nonumber\\[0.5em]
    \overline{S}_2 = \sin{2\gamma_{23}} - \sin{2\gamma_{12}}, \quad \overline{C}_2 = \cos{2\gamma_{23}} - \cos{2\gamma_{12}} \nonumber \\[0.5em]
    \overline{S}_3 = \sin{2\gamma_{31}} - \sin{2\gamma_{23}}, \quad \overline{C}_3 = \cos{2\gamma_{31}} - \cos{2\gamma_{23}} 
\end{align}
The matrix $\bB_{R}\transpose$ in equation \ref{eq:BR1}, \ref{eq:BR2} and \ref{eq:BR3} can be rewritten as:
\begin{align}\label{eq:BRS}
    \bB_{R_1}\transpose = \frac{1}{2} \, \overline{S}_1\left(\bB_{M_y^1}\transpose -\bB_{M_x^1}\transpose\right) + \overline{C}_1\bB_{M_{xy}^1}\transpose \nonumber\\[0.5em]
    \bB_{R_2}\transpose = \frac{1}{2} \, \overline{S}_2\left(\bB_{M_y^2}\transpose -\bB_{M_x^2}\transpose\right) + \overline{C}_2\bB_{M_{xy}^2}\transpose \nonumber\\[0.5em]
    \bB_{R_3}\transpose = \frac{1}{2} \, \overline{S}_3\left(\bB_{M_y^3}\transpose -\bB_{M_x^3}\transpose\right) + \overline{C}_3\bB_{M_{xy}^3}\transpose   
\end{align}

Next, we calculate $V_n$ according to equations \ref{eq:moment1-2} to \ref{eq:moment2-2}. Plugging equations \ref{eq:moment2-1} and \ref{eq:moment2-2} into \ref{eq:moment1-2}, we get:
\begin{align}\label{eq:appendix_Vn}
    V_n =\frac{\partial M_{n}}{\partial n} + 2 \frac{\partial M_{ns}}{\partial s}
\end{align}
Plugging equation \ref{eq:directional derivatives} into equation \ref{eq:appendix_Vn}, we get:
\begin{align}\label{eq:appendix_Vn_dxdy}
    V_n = \cos\gamma\frac{\partial M_n}{\partial x} + \sin\gamma\frac{\partial M_n}{\partial y} - 2\sin\gamma\frac{\partial M_{ns}}{\partial x} + 2\cos\gamma\frac{\partial M_{ns}}{\partial y}
\end{align}
The terms on the RHS above require eventually the evaluation of $\frac{\partial M_n}{\partial x}$, $\frac{\partial M_n}{\partial y}$, $\frac{\partial M_{ns}}{\partial x}$, and $\frac{\partial M_{ns}}{\partial y}$.

For $V_n^{12}$ item, the derivation from equation \ref{eq:Mn_12} is:
\begin{align}
    \frac{\partial M_{n}^{12}}{\partial x}  = \cos^2{\gamma_{12}}\frac{\partial \bB_{M_x^1}\transpose}{\partial x}\balpha + \sin^2{\gamma_{12}}\frac{\partial \bB_{M_y^1}\transpose}{\partial x}\balpha + \sin{2\gamma_{12}}\frac{\partial \bB_{M_{xy}^1}\transpose}{\partial x}\balpha\\[0.5em]
    \frac{\partial M_{n}^{12}}{\partial y}  = \cos^2{\gamma_{12}}\frac{\partial \bB_{M_x^1}\transpose}{\partial y}\balpha + \sin^2{\gamma_{12}}\frac{\partial \bB_{M_y^1}\transpose}{\partial y}\balpha + \sin{2\gamma_{12}}\frac{\partial \bB_{M_{xy}^1}\transpose}{\partial y}\balpha 
\end{align}
where
\begin{align}
    \frac{\partial \bB_{M_x^1}\transpose}{\partial x} &= \begin{bmatrix}
    0, \ 0 , \ 0 , \ -6D_{11}, \ -4D_{16} , \ -2D_{12} , \ 0 
    \end{bmatrix} \\[0.5em]
    \frac{\partial \bB_{M_y^1}\transpose}{\partial x} &= \begin{bmatrix}
    0 , \ 0 , \ 0 , \  -6D_{12} , \ -4D_{26} , \ -2D_{22} , \ 0
    \end{bmatrix}  \\[0.5em]
    \frac{\partial \bB_{M_{xy}^1}\transpose}{\partial x} &= \begin{bmatrix}
    0 , \ 0 , \ 0 , \ -6D_{16} , \ -4D_{66} , \ -2D_{26} , \ 0
    \end{bmatrix} 
\end{align}
And 
\begin{align}
    \frac{\partial \bB_{M_x^1}\transpose}{\partial y} &= \begin{bmatrix}
    0, \  0, \ 0, \ 0 , \ -2D_{11} , \ -4D_{16} , \ -6D_{12} 
    \end{bmatrix} \\[0.5em]
    \frac{\partial \bB_{M_y^1}\transpose}{\partial y} &= \begin{bmatrix}
    0, \ 0, \ 0, \  0, \ -2D_{12}, \ -4D_{26}, \ -6D_{22}
    \end{bmatrix}  \\[0.5em]
    \frac{\partial \bB_{M_{xy}^1}\transpose}{\partial y} &= \begin{bmatrix}
    0, \ 0, \ 0, \ 0,\ -2D_{16} , \ -4D_{66}, \ -6D_{26}
    \end{bmatrix}  
\end{align}
Thus, the matrix $\bB_{M_{n,x}^{12}}\transpose$ and $\bB_{M_{n,y}^{12}}\transpose$ are:
\begin{align}
    \bB_{M_{n,x}^{12}}\transpose = \cos^2{\gamma_{12}}\frac{\partial \bB_{M_x^1}\transpose}{\partial x} + \sin^2{\gamma_{12}}\frac{\partial \bB_{M_y^1}\transpose}{\partial x} + \sin{2\gamma_{12}}\frac{\partial \bB_{M_{xy}^1}\transpose}{\partial x}\\[0.5em]    
    \bB_{M_{n,y}^{12}}\transpose = \cos^2{\gamma_{12}}\frac{\partial \bB_{M_x^1}\transpose}{\partial y} + \sin^2{\gamma_{12}}\frac{\partial \bB_{M_y^1}\transpose}{\partial y} + \sin{2\gamma_{12}}\frac{\partial \bB_{M_{xy}^1}\transpose}{\partial y} 
\end{align}

Using equation \ref{eq:moment2-1}, the derivative of moment $M_{ns}$ on side 1-2 is:
\begin{align}
    \frac{\partial M_{ns}^{12}}{\partial x} = \left[\frac{1}{2}\left(\frac{\partial \bB_{M_y^1}\transpose}{\partial x} - \frac{\partial \bB_{M_x^1}\transpose}{\partial x}\right)\sin 2\gamma_{12} + \frac{\partial \bB_{M_{xy}^1}\transpose }{\partial x} \cos 2\gamma_{12}\right]\,\balpha \\[0.5em]
    \frac{\partial M_{ns}^{12}}{\partial y} = \left[\frac{1}{2}\left(\frac{\partial \bB_{M_y^1}\transpose}{\partial y} - \frac{\partial \bB_{M_x^1}\transpose}{\partial y}\right)\sin 2\gamma_{12} + \frac{\partial \bB_{M_{xy}^1}\transpose }{\partial y} \cos 2\gamma_{12}\right]\,\balpha    
\end{align}
Thus, the matrices $\bB_{M_{ns,x}^{12}}\transpose$ and $\bB_{M_{ns,y}^{12}}\transpose$ are:
\begin{align}
    \bB_{M_{ns,x}^{12}}\transpose = \frac{1}{2}\sin 2\gamma_{12}\left(\frac{\partial \bB_{M_y^1}\transpose}{\partial x} - \frac{\partial \bB_{M_x^1}\transpose}{\partial x}\right) + \cos 2\gamma_{12}\frac{\partial \bB_{M_{xy}^1}\transpose }{\partial x} \\[0.5em]
    \bB_{M_{ns,y}^{12}}\transpose = \frac{1}{2}\sin 2\gamma_{12}\left(\frac{\partial \bB_{M_y^1}\transpose}{\partial y} - \frac{\partial \bB_{M_x^1}\transpose}{\partial y}\right) + \cos 2\gamma_{12}\frac{\partial \bB_{M_{xy}^1}\transpose }{\partial y}     
\end{align}

Finally, using equation \ref{eq:appendix_Vn_dxdy}, $\bB$ matrix for the Kirchhoff shear force $V_n$ on side 1-2 is:
\begin{align}
    \bB_{V_n^{12}}\transpose = &\cos{\gamma_{12}} \bB_{M_{n,x}^{12}}\transpose + \sin{\gamma_{12}}\bB_{M_{n,y}^{12}}\transpose - 2\sin{\gamma_{12}}\bB_{M_{ns,x}^{12}}\transpose + 2\cos{\gamma_{12}}\bB_{M_{ns,y}^{12}}\transpose
\end{align}

Similarly, $\bB_{V_n}$ matrix on side 2-3 is:
\begin{align}
    \bB_{M_{n,x}^{23}}\transpose = \cos^2{\gamma_{23}}\frac{\partial \bB_{M_x^2}\transpose}{\partial x} + \sin^2{\gamma_{23}}\frac{\partial \bB_{M_y^2}\transpose}{\partial x} + \sin{2\gamma_{23}}\frac{\partial \bB_{M_{xy}^1}\transpose}{\partial x}\\[0.5em]    
    \bB_{M_{n,y}^{23}}\transpose = \cos^2{\gamma_{23}}\frac{\partial \bB_{M_x^2}\transpose}{\partial y} + \sin^2{\gamma_{23}}\frac{\partial \bB_{M_y^2}\transpose}{\partial y} + \sin{2\gamma_{23}}\frac{\partial \bB_{M_{xy}^2}\transpose}{\partial y} 
\end{align}
and
\begin{align}
    \bB_{M_{ns,x}^{23}}\transpose = \frac{1}{2}\sin 2\gamma_{23}\left(\frac{\partial \bB_{M_y^2}\transpose}{\partial x} - \frac{\partial \bB_{M_x^2}\transpose}{\partial x}\right) + \cos 2\gamma_{23}\frac{\partial \bB_{M_{xy}^2}\transpose }{\partial x} \\[0.5em]
    \bB_{M_{ns,y}^{23}}\transpose = \frac{1}{2}\sin 2\gamma_{23}\left(\frac{\partial \bB_{M_y^2}\transpose}{\partial y} - \frac{\partial \bB_{M_x^2}\transpose}{\partial y}\right) + \cos 2\gamma_{23}\frac{\partial \bB_{M_{xy}^2}\transpose }{\partial y}    
\end{align}

The $\bB$ matrix for $V_n^{23}$ is:
\begin{align}
    \bB_{V_n^{23}}\transpose = &\cos{\gamma_{23}} \bB_{M_{n,x}^{23}}\transpose + \sin{\gamma_{23}}\bB_{M_{n,y}^{23}}\transpose - 2\sin{\gamma_{23}}\bB_{M_{ns,x}^{23}}\transpose + 2\cos{\gamma_{23}}\bB_{M_{ns,y}^{23}}\transpose
\end{align}

On side 3-1:
\begin{align}\label{eq:B_M_n,x31}
    \bB_{M_{n,x}^{31}}\transpose = \cos^2{\gamma_{31}}\frac{\partial \bB_{M_x^3}\transpose}{\partial x} + \sin^2{\gamma_{31}}\frac{\partial \bB_{M_y^3}\transpose}{\partial x} + \sin{2\gamma_{31}}\frac{\partial \bB_{M_{xy}^3}\transpose}{\partial x}\\[0.5em]    
    \bB_{M_{n,y}^{31}}\transpose = \cos^2{\gamma_{31}}\frac{\partial \bB_{M_x^3}\transpose}{\partial y} + \sin^2{\gamma_{31}}\frac{\partial \bB_{M_y^3}\transpose}{\partial y} + \sin{2\gamma_{31}}\frac{\partial \bB_{M_{xy}^3}\transpose}{\partial y} 
\end{align}
and
\begin{align}
    \bB_{M_{ns,x}^{31}}\transpose = \frac{1}{2}\sin 2\gamma_{31}\left(\frac{\partial \bB_{M_y^3}\transpose}{\partial x} - \frac{\partial \bB_{M_x^3}\transpose}{\partial x}\right) + \cos 2\gamma_{31}\frac{\partial \bB_{M_{xy}^3}\transpose }{\partial x} \\[0.5em]
    \bB_{M_{ns,y}^{31}}\transpose = \frac{1}{2}\sin 2\gamma_{31}\left(\frac{\partial \bB_{M_y^3}\transpose}{\partial y} - \frac{\partial \bB_{M_x^3}\transpose}{\partial y}\right) + \cos 2\gamma_{31}\frac{\partial \bB_{M_{xy}^3}\transpose }{\partial y}   
\end{align}

The $\bB$ matrix for $V_n^{31}$ is:
\begin{align}
    \bB_{V_n^{31}}\transpose = &\cos{\gamma_{31}} \bB_{M_{n,x}^{31}}\transpose + \sin{\gamma_{23}}\bB_{M_{n,y}^{31}}\transpose \nonumber - 2\sin{\gamma_{23}}\bB_{M_{ns,x}^{31}}\transpose + 2\cos{\gamma_{23}}\bB_{M_{ns,y}^{31}}\transpose        
\end{align}
Therefore, the calculations of $\bB$ matrices for $V_n$ on sides 23 and 31 can be simplified through the use of the pre-calculated matrices for side 12. 

Now, all the components of $\bB$ matrix have been derived. The assembled (7 x 12) matrix $\bB$ is:
\begin{align}
    \bB  = 
    [&\bB_{R_1} ,
    \bB_{R_2} ,
    \bB_{R_3} ,
    \bB_{V_{n}^{12}} ,
    \bB_{V_{n}^{23}} ,
    \bB_{V_{n}^{31}} , 
    \bB_{M_{n}^{12}} ,
    \bB_{M_{n}^{21}} ,
    \bB_{M_{n}^{23}} ,
    \bB_{M_{n}^{32}} ,
    \bB_{M_{n}^{31}} ,
    \bB_{M_{n}^{13}} ]
\end{align}

\section{Analytical solution of SLB benchmark}
\setcounter{figure}{0} 
\label{sec:analytical-benchmarks}
\begin{figure}[h]
	\centering
	\includegraphics[width=0.75\linewidth]{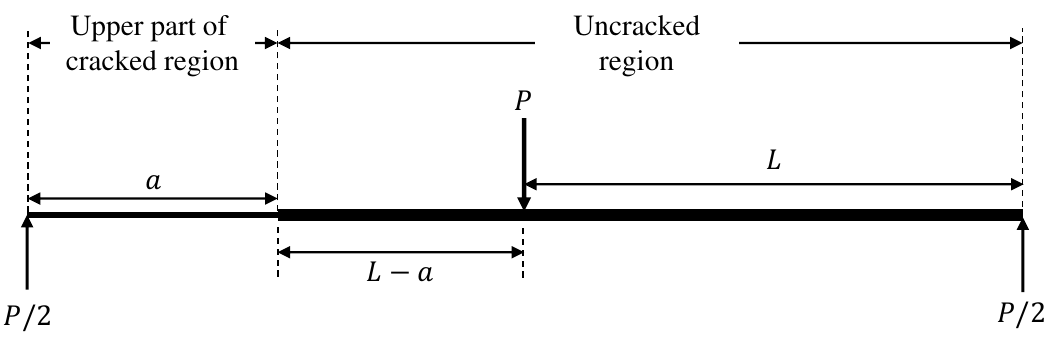}
	\caption{Analytical solution of SLB model}
	\label{fig: SLB_analytical}
\end{figure}
Referring to Figure~\ref{fig: SLB_analytical}, the analytical load-displacement curve consists of three parts. The first part, OB, is the linear elastic stage with the initial crack length $a_0$. The second part comes from the curve ABC which represents the load-displacement response during delamination propagation when $a<L$ (the half length of the SLB model). When $a>L$, we can obtain the last curve DFE. The full analytical curve would then be OBFE. The derivation process of each curve is shown one by one in the following content.

\begin{figure}[h]
	\centering
	\includegraphics[width=0.8\linewidth]{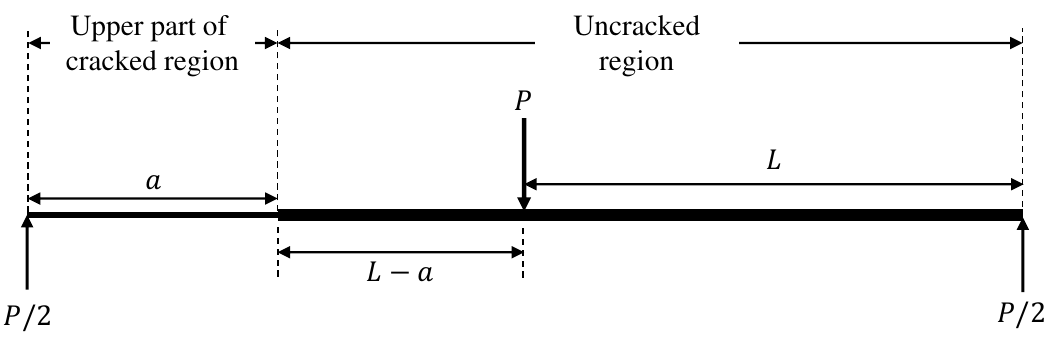}
	\caption{Beam model of SLB specimen ($a<L$)}
	\label{fig: SLB_analytical a<L}
\end{figure}

When the crack length $a$ is smaller than $L$, the analytical derivation of the SLB benchmark follows from the work of Davidson et el. \cite{davidson1995three}. Referring to the SLB model in Figure~\ref{fig:SLB}, it is evident that the bending stiffness in the cracked region comes entirely from the part above the crack, which is midplane symmetrical. The uncracked region is midplane anti-symmetric. 

The SLB model ($a<L$) is simplified into a 2D beam model, as shown in Figure \ref{fig: SLB_analytical a<L}. The moment-curvature relationship for the SLB problem can be expressed as:
\begin{align}\label{eq:moment-curvature}
    M = -b \, D \, \frac{\dd^2 w}{\dd x^2} = b \, D \,\kappa_x
\end{align}
where $M$ is the moment on a face along the x-axis, $b$ is the width of the specimen, $w$ is the deflection along the $z$ direction. And $D$ is the effective bending rigidity per unit width. Because the values of $D$ in the uncracked region and cracked regions are different, $D_0$ represents the effective bending rigidity of the uncracked region, and $D_1$ represents the effective bending rigidity of the cracked region to distinguish them.

We assume that the classical laminate theory is applicable to the SLB problem. In matrix form, the constitutive equation can be written as:
\begin{align}\label{eq:ADBconstitutive}
    \begin{bmatrix}
    N_x \\[0.5em]
    N_y \\[0.5em]
    N_{xy} \\[0.5em]
    -- \\[0.5em]
    M_x \\[0.5em]
    M_y \\[0.5em]
    M_{xy}
    \end{bmatrix} = 
    \begin{bmatrix}
    A_{11} & A_{12} & A_{16} & | & B_{11} & B_{12} & B_{16} \\[0.5em]
    A_{12} & A_{22} & A_{26} & | & B_{12} & B_{22} & B_{26} \\[0.5em]
    A_{16} & A_{26} & A_{66} & | & B_{16} & B_{26} & B_{66} \\[0.5em]
    -- & -- & -- & | & -- & -- & -- \\[0.5em]
    B_{11} & B_{12} & B_{16} & | & D_{11} & D_{12} & D_{16} \\[0.5em]
    B_{12} & B_{22} & B_{26} & | & D_{12} & D_{22} & D_{26} \\[0.5em]
    B_{16} & B_{26} & B_{66} & | & D_{16} & D_{26} & D_{66}
    \end{bmatrix} 
    \begin{bmatrix}
    \varepsilon_x^0 \\[0.5em]
    \varepsilon_y^0 \\[0.5em]
    \gamma_{xy}^0 \\[0.5em]
    -- \\[0.5em]
    \kappa_x \\[0.5em]
    \kappa_y \\[0.5em]
    \kappa_{xy}
    \end{bmatrix}
\end{align}
where $\varepsilon_x^0, \, \varepsilon_y^0, \, \gamma_{xy}^0$ are the membrane strains of the neutral plane, and $ \kappa_x, \, \kappa_y, \, \kappa_{xy} $ are the curvatures, derived from the out-of-plane displacement $w$:
\begin{align}
    \kappa_x=-\frac{\partial^2 w}{\partial x^2}, \quad \kappa_y=-\frac{\partial^2 w}{\partial y^2}, \quad \kappa_{xy}=-\frac{\partial^2 w}{\partial x \partial y}
\end{align}

The inverted form of the constitutive relationship is:
\begin{align}\label{eq:ADBconstitutive_inverse}
    \begin{bmatrix}
    \varepsilon_x^0 \\[0.5em]
    \varepsilon_y^0 \\[0.5em]
    \gamma_{xy}^0 \\[0.5em]
    -- \\[0.5em]
    \kappa_x \\[0.5em]
    \kappa_y \\[0.5em]
    \kappa_{xy} 
    \end{bmatrix}  
 = 
    \begin{bmatrix}
    \alpha_{11} & \alpha_{12} & \alpha_{16} & | & \beta_{11} & \beta_{12} & \beta_{16} \\[0.5em]
    \alpha_{12} & \alpha_{22} & \alpha_{26} & | & \beta_{12} & \beta_{22} & \beta_{26} \\[0.5em]
    \alpha_{16} & \alpha_{26} & \alpha_{66} & | & \beta_{16} & \beta_{26} & \beta_{66} \\[0.5em]
    -- & -- & -- & | & -- & -- & -- \\[0.5em]
    \beta_{11} & \beta_{12} & \beta_{16} & | & \delta_{11} & \delta_{12} & \delta_{16} \\[0.5em]
    \beta_{12} & \beta_{22} & \beta_{26} & | & \delta_{12} & \delta_{22} & \delta_{26} \\[0.5em]
    \beta_{16} & \beta_{26} & \beta_{66} & | & \delta_{16} & \delta_{26} & \delta_{66}
    \end{bmatrix} 
     \begin{bmatrix}
    N_x \\[0.5em]
    N_y \\[0.5em]
    N_{xy} \\[0.5em]
    -- \\[0.5em]
    M_x \\[0.5em]
    M_y \\[0.5em]
    M_{xy}
    \end{bmatrix} 
\end{align}

If the plate constraint condition is ``generalized plane stress", $N_y$, $N_{xy}$, $M_y$, and $M_{xy}$ are zero. When considering $N_x = 0$, equation \ref{eq:ADBconstitutive_inverse} gives:
\begin{align}\label{eq:D_plainstress}
    D = \frac{1}{\delta_{11}}
\end{align}
For the plane strain condition ($\epsilon_y^0 = \gamma_{xy}^0 =  \kappa_y = \kappa_{xy} = 0$), equation \ref{eq:ADBconstitutive} gives:
\begin{align}\label{eq:D_plainstrain}
    D = D_{11}
\end{align} 

The expression of strain energy is:
\begin{align}\label{eq:Strain energy (Beam)_Re}
    U = \int_0^{2L} \frac{M^2\mathrm{d}x}{2bD}  
\end{align}

From Figure~\ref{fig: SLB_analytical a<L}, the boundary conditions of the SLB model can be considered similar to three-point bending. So we can get the moment along the beam as:
\begin{align}\label{eq:moment of beam}
    M = \begin{cases}
 -\dfrac{Px}{2}, &  0 \leq x \leq L \\[1em]
 \dfrac{Px}{2}-PL ,& L \leq x \leq 2L \\
\end{cases}
\end{align}
Since the cross-section of the upper part of the cracked region $(0 \leq x \leq a)$ is different from that of the uncracked region, the moment $M$ is further divided, and the expression for strain energy is the sum of three parts:
\begin{align}\label{eq:sum of U}
    \left.U\right\vert_{(a<L)} = \int_{0}^{a} \frac{M^2\mathrm{d}x}{2bD_1}  + \int_{a}^{L-a} \frac{M^2\mathrm{d}x}{2bD_0} + \int_{L}^{2L} \frac{M^2\mathrm{d}x}{2bD_0} 
\end{align}
Substituting equation \ref{eq:moment of beam} into equation \ref{eq:sum of U}, we can get:
\begin{align}
\left.U\right\vert_{(a<L)}= \int_{0}^{a}  \frac{\left({-\frac{Px}{2}}\right)^2\mathrm{d}x}{2bD_1}  + \int_{a}^{L} \frac{\left({-\frac{Px}{2}}\right)^2\mathrm{d}x}{2bD_0} + \int_{L}^{2L} \frac{\left({\frac{Px}{2}-PL}\right)^2\mathrm{d}x}{2bD_0} 
\end{align}

Based on the expression of strain energy, the displacement $\delta$ can be calculated by Castigliano's second theorem:
\begin{align}\label{eq:appendix D: Delta}
    \left.\delta \right\vert_{(a<L)}&= \frac{\partial{\left.U\right\vert_{(a<L)}}}{\partial{P}}  = \int_{0}^{a}  \frac{Px^2\mathrm{d}x}{4bD_1}  + \int_{a}^{L} \frac{Px^2\mathrm{d}x}{4bD_0} + \int_{L}^{2L} \frac{\left({Px^2-4PLx+4PL^2}\right)\mathrm{d}x}{4bD_0} \nonumber\\[0.5em]
           &= \frac{Pa^3}{12bD_1} + \frac{2PL^3-Pa^3}{12bD_0} 
\end{align}
The above equation can be used to plot the $P-\delta$ curve for a certain crack length $a$, which for the case of $a=a_0$ gives the first part of the analytical curve, i.e., OB.
The compliance $C$ during the elastic region is defined as the center deflection $\delta$ divided by the center load $P$:
\begin{align}\label{eq: compliance of SLB}
    \left.C\right\vert_{(a<L)} &= \frac{\left.\delta \right\vert_{(a<L)}}{P} = \frac{2L^3 + a^3(R-1)}{12bD_0} 
\end{align}
where $R$ is the ratio of the bending rigidity of the uncracked region to that of the cracked region:
\begin{align}\label{eq:R appendix}
    R = D_0/D_1
\end{align}

The relationship between the critical energy release rate $G_c$ and the derivative of compliance $C$ with respect to the crack length $a$ is:
\begin{align}\label{eq: ERR-derivative C}
    G_c = \frac{P^2}{2b} \, \frac{\partial  C }{\partial a}
\end{align}
Substituting equation \ref{eq: compliance of SLB} into equation \ref{eq: ERR-derivative C}, we can obtain:
\begin{align}\label{eq: ERR of SLB}
    G_c = \frac{P^2a^2(R-1)}{8b^2D_0} 
\end{align}
Substituting equation \ref{eq: compliance of SLB} to equation \ref{eq: ERR of SLB}, the critical energy release rate $G_c$ can also be obtained by the critical load and deflection:
\begin{align}\label{eq: ERR of SLB -2}
    G_c = \frac{3Pa^2\left.\delta \right\vert_{(a<L)} }{2b} \, \frac{(R-1)}{[2L^3+a^3(R-1)]}
\end{align}

The critical energy release rate $G_c$ in equation \ref{eq: ERR of SLB -2} is the value in mixed mode. It can be calculated by:
\begin{align}\label{eq:appendix Gc with B}
    G_\mathrm{c} = G_\mathrm{Ic} + (G_\mathrm{IIc}-G_\mathrm{Ic})B^{\eta}
\end{align}
where $G_\mathrm{Ic}$ and $G_\mathrm{IIc}$ are the critical energy release rates in pure mode I and mode II, respectively. The mixed mode ratio $B$ in the SLB model is 0.4 \cite{krueger2015summary}.
Equation~\ref{eq: ERR of SLB} can be used to calculate the critical load $P$ at a certain crack length $a$. It can also be used to express $a$ in terms of $P$ for a fixed $G_c$, which can be plugged into equation~\ref{eq: ERR of SLB -2} to obtain the relationship between $\delta$ and $P$ for a fixed $G_c$ for the case of $a<L$, hence giving the curve ABC.

 \begin{figure}[htbp]
	\centering
	\includegraphics[width=0.8\linewidth]{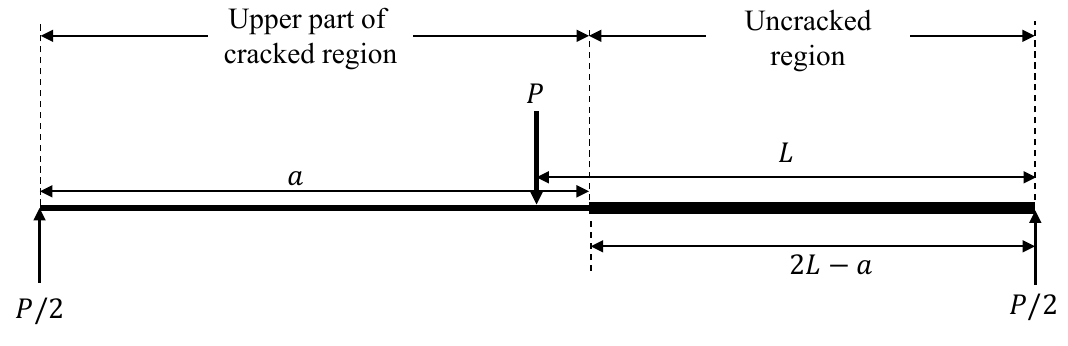}
	\caption{Beam model of SLB specimen ($a\geq L$)}
	\label{fig: SLB_analytical a>L}
\end{figure}

When the crack length $a$ is greater than or equal to $L$, the analytical derivation of the SLB benchmark can follow the same derivation process as above. The SLB model ($a\geq L$) is simplified in Figure \ref{fig: SLB_analytical a>L}. Based on Figure ~\ref{fig: SLB_analytical a>L}, the expression for strain energy is rewritten as:
\begin{align}\label{eq:sum of U a>L}
   \left.U\right\vert_{(a\geq L)} = \int_{0}^{L} \frac{M^2\mathrm{d}x}{2bD_1}  + \int_{L}^{a} \frac{M^2\mathrm{d}x}{2bD_1} + \int_{a}^{2L} \frac{M^2\mathrm{d}x}{2bD_0} 
\end{align}

Since the boundary conditions have not changed, the expression of the moment is consistent with equation \ref{eq:moment of beam}. Substituting equation \ref{eq:moment of beam} into equation \ref{eq:sum of U a>L}, we can obtain:
\begin{align}\label{eq:sum of U a>L with M}
 \left.U\right\vert_{(a\geq L)} = &\int_{0}^{L}  \frac{P^2x^2\mathrm{d}x}{8bD_1} 
+ \int_{L}^{a} \frac{\left(P^2x^2-4P^2Lx+4P^2L^2\right)\mathrm{d}x}{8bD_1} \nonumber\\[0.5em]
&+ \int_{a}^{2L} \frac{\left(P^2x^2-4P^2Lx+4P^2L^2\right)\mathrm{d}x}{8bD_0} 
\end{align}

The displacement $\delta$ at the middle can also be found by Castigliano's second theorem:
\begin{align}
\left.\delta\right\vert_{(a\geq L)} 
= -\frac{PL^3}{2bD_1} + \frac{Pa^3-6PLa^2+12PL^2a}{12bD_1} + \frac{8PL^3}{12bD_0} - \frac{Pa^3-6PLa^2+12PL^2a}{12bD_0}
\end{align}
Introducing ratio $R$ in equation \ref{eq:R appendix}, the above equation can be rewritten as:
\begin{align}\label{eq:appendix D: Delta a>L}
    \left.\delta\right\vert_{(a\geq L)}=  -\frac{(6R-8)PL^3}{12bD_0}+  \frac{(R-1)(Pa^3-6PLa^2+12PL^2a)}{12bD_0}
\end{align}

The expression of compliance $C$ under the condition $a \geq L$ is:
\begin{align}\label{eq: compliance of SLB a>L}
    \left.C\right\vert_{(a \geq L)} &= \frac{\left.\delta \right\vert_{(a \geq L)}}{P} = -\frac{(6R-8)L^3}{12bD_0}+  \frac{(R-1)(a^3-6La^2+12L^2a)}{12bD_0}
\end{align}
Substituting equation \ref{eq: compliance of SLB a>L} into equation \ref{eq: ERR-derivative C}, we have:
\begin{align}\label{eq: ERR of SLB a>L}
    G_c = \frac{P^2(R-1)(3a^2-12aL+12L^2)}{24b^2D_0} 
\end{align}
where $G_c$ is the critical energy release rate given by equation~\ref{eq:appendix Gc with B}. Using equations~\ref{eq:appendix D: Delta a>L} and \ref{eq: ERR of SLB a>L}, one can then vary $a$ to plot the third part of the analytical curve DE.





\end{document}

%% file: commands.tex
\newcommand{\I}{\rm{I}}
\newcommand{\II}{\rm{II}}
\newcommand{\lce}{l}
\newcommand{\beam}{\rm{beam}}
\newcommand{\btm}{\rm{bot}} 
\newcommand{\CE}{\rm{CE}}
\newcommand{\CR}{\rm{CR}}
\newcommand{\dd}{\mathrm{d}} 
\newcommand{\external}{\rm{ext}}
\newcommand{\h}{\rm{h}}
\newcommand{\internal}{\rm{int}}
\newcommand{\len}{\rm{\mathcal{l}}}
\newcommand{\truss}{\rm{truss}}
\newcommand{\tp}{\rm{top}} 
\newcommand{\transpose}{^{\rm{T}}}
\newcommand{\invtranspose}{^{\rm{-T}}}
\newcommand{\xl}{{\hat{x}}}
\newcommand{\yl}{{\hat{y}}}
\newcommand{\inv}{\makebox[0pt][l]{$^{-1}$}}
\newcommand{\DeltaI}{\Delta_{\text{I}}}
\newcommand{\DeltaII}{\Delta_{\text{II}}}
\newcommand{\DeltaIII}{\Delta_{\text{III}}}
\newcommand{\wtopce}{w^{\text{topCE}}}
\newcommand{\wbotce}{w^{\text{botCE}}}
\newcommand{\utopce}{u^{\text{topCE}}}
\newcommand{\ubotce}{u^{\text{botCE}}}
\newcommand{\vtopce}{v^{\text{topCE}}}
\newcommand{\vbotce}{v^{\text{botCE}}}
\newcommand{\wtop}{w^{\text{top}}}
\newcommand{\wbot}{w^{\text{bot}}}
\newcommand{\utop}{u^{\text{top}}}
\newcommand{\ubot}{u^{\text{bot}}}
\newcommand{\vttop}{v^{\text{top}}}
\newcommand{\vbot}{v^{\text{bot}}}
\newcommand{\htop}{h^{\text{top}}}
\newcommand{\hbot}{h^{\text{bot}}}
\newcommand{\thetatop}{\theta^{\text{top}}}
\newcommand{\thetabot}{\theta^{\text{bot}}}

\newcommand{\ba}{\boldsymbol{\mathrm{a}}}
\newcommand{\bA}{\boldsymbol{\mathrm{A}}}
\newcommand{\bb}{\boldsymbol{\mathrm{b}}}
\newcommand{\bB}{\boldsymbol{\mathrm{B}}}
\newcommand{\bBN}{\bB_{\text{N}}}
\newcommand{\bC}{\boldsymbol{\mathrm{C}}}
\newcommand{\D}{\boldsymbol{\mathrm{D}}}
\newcommand{\Dce}{\D^{\CE}}
\newcommand{\bE}{\boldsymbol{\mathrm{E}}}
\newcommand{\bF}{\boldsymbol{\mathrm{F}}}
\newcommand{\f}{\boldsymbol{\mathrm{f}}} 
\newcommand{\fext}{\f_{\external}}
\newcommand{\fint}{\f_{\internal}}
\newcommand{\fhatext}{\hat{\f}_{\external}}
\newcommand{\bH}{\boldsymbol{\mathrm{H}}}
\newcommand{\bI}{\boldsymbol{\mathrm{I}}}
\newcommand{\bJ}{\boldsymbol{\mathrm{J}}}
\newcommand{\K}{\boldsymbol{\mathrm{K}}}
\newcommand{\Kmat}{\K_{\mathrm{mat}}}
\newcommand{\Kgeo}{\K_{\mathrm{geo}}}
\newcommand{\bM}{\boldsymbol{\mathrm{M}}}
\newcommand{\bn}{\boldsymbol{\mathrm{n}}}
\newcommand{\bN}{\boldsymbol{\mathrm{N}}}
\newcommand{\bp}{\boldsymbol{\mathrm{p}}}
\newcommand{\bq}{\boldsymbol{\mathrm{q}}}
\newcommand{\bqce}{\bq^{\CE}}
\newcommand{\bQ}{\boldsymbol{\mathrm{Q}}}
\newcommand{\bU}{\boldsymbol{\mathrm{U}}}
\newcommand{\bR}{\boldsymbol{\mathrm{R}}}
\newcommand{\bS}{\boldsymbol{\mathrm{S}}}
\newcommand{\bt}{\boldsymbol{\mathrm{t}}}
\newcommand{\bT}{\boldsymbol{\mathrm{T}}}
\newcommand{\bu}{\boldsymbol{\mathrm{u}}}
\newcommand{\bv}{\boldsymbol{\mathrm{v}}}
\newcommand{\bw}{\boldsymbol{\mathrm{w}}}
\newcommand{\bW}{\boldsymbol{\mathrm{W}}}
\newcommand{\bX}{\boldsymbol{\mathrm{X}}}
\newcommand{\bx}{\boldsymbol{\mathrm{x}}}

\newcommand{\balpha}{\boldsymbol{\mathrm{\alpha}}}
\newcommand{\bDelta}{\boldsymbol{\mathrm{\Delta}}}
\newcommand{\beps}{\boldsymbol{\mathrm{\epsilon}}}
\newcommand{\bgamma}{\boldsymbol{\mathrm{\gamma}}}
\newcommand{\bPhi}{\boldsymbol{\Phi}}
\newcommand{\bsig}{\boldsymbol{\mathrm{\sigma}}}
\newcommand{\btau}{\boldsymbol{\mathrm{\tau}}}
\newcommand{\bxi}{\boldsymbol{\mathrm{\xi}}}

\newcommand{\minus}{\scalebox{0.75}[1.0]{$-$}}